\documentclass{aa}  

\usepackage{natbib}
\bibpunct{(}{)}{;}{a}{}{,}
\usepackage{graphicx}
\usepackage[varg]{txfonts}
\usepackage{hyperref}
\hypersetup{
     colorlinks   = true,
     citecolor    = blue
}
\usepackage{units}

\begin{document} 
\defcitealias{Bre:16}{BFS+}

\title{Numerical studies on the link between radioisotopic signatures on Earth and the formation of the Local Bubble}
\subtitle{I.~\element[][60]{Fe} transport to the solar system by turbulent mixing of ejecta from nearby supernovae into a locally homogeneous interstellar medium}

\author{M.~M. Schulreich\inst{\ref{inst1}} \and D.~Breitschwerdt\inst{\ref{inst1}} \and J.~Feige\inst{\ref{inst1}} \and C.~Dettbarn\inst{\ref{inst2}}}
\institute{Zentrum f\"ur Astronomie und Astrophysik, Technische Universit\"at Berlin, Hardenbergstra\ss e~36, 10623 Berlin, Germany\\\email{schulreich@astro.physik.tu-berlin.de}\label{inst1} \and Astronomisches Rechen-Institut, Zentrum f\"ur Astronomie der Universit\"at Heidelberg, M\"onchhofstra\ss e 12--14, 69120 Heidelberg, Germany\label{inst2}}

\date{Received / Accepted}

\titlerunning{Numerical studies on the link between radioisotopic signatures on Earth and the formation of the LB. I}
\authorrunning{M.~M.~Schulreich et al.}

\abstract{The discovery of radionuclides like \element[][60]{Fe} with half-lives of million years in deep-sea crusts and sediments offers the unique possibility to date and locate nearby supernovae.} 
{We want to quantitatively establish that the \element[][60]{Fe} enhancement is the result of several supernovae which are also responsible for the formation of the Local Bubble, our Galactic habitat.}
{We performed three-dimensional hydrodynamic adaptive mesh refinement simulations (with resolutions down to subparsec scale) of the Local Bubble and the neighbouring Loop~I superbubble in different homogeneous, self-gravitating environments. For setting up the Local and Loop~I superbubble, we took into account the time sequence and locations of the generating core-collapse supernova explosions, which were derived from the mass spectrum of the perished members of certain stellar moving groups. The release of \element[][60]{Fe} and its subsequent turbulent mixing process inside the superbubble cavities was followed via passive scalars, where the yields of the decaying radioisotope were adjusted according to recent stellar evolution calculations.}
{The models are able to reproduce both the timing and the intensity of the \element[][60]{Fe} excess observed with rather high precision, provided that the external density does not exceed $\unit[0.3]{cm^{-3}}$ on average. Thus the two best-fit models presented here were obtained with background media mimicking the classical warm ionised and warm neutral medium. We also found that \element[][60]{Fe} (which is condensed onto dust grains) can be delivered to Earth via two physical mechanisms: either through individual fast-paced supernova blast waves, which cross the Earth's orbit sometimes even twice as a result of reflection from the Local Bubble's outer shell, or, alternatively, through the supershell of the Local Bubble itself, injecting the \element[][60]{Fe} content of all previous supernovae at once, but over a longer time range.} 
{}
\keywords{ISM: bubbles -- ISM: supernova remnants -- ISM: abundances -- hydrodynamics -- methods: numerical }
\maketitle
%

\section{Introduction}
\label{ch:intro}
Our solar system resides in a low-density region of the interstellar medium (ISM), partially filled with hot, soft X-ray emitting plasma, known as the Local Bubble (LB). While the existence of this superbubble (SB) is widely accepted today, not least because it was found to be responsible for about 60\,\% of the $1/4$-keV flux in the Galactic plane \citep{Gal:14}, its exact origin is still a matter of ongoing debate. The LB extends roughly $\unit[200]{pc}$ in the Galactic plane, and $\unit[600]{pc}$ perpendicular to it, with an inclination of about $20^\circ$ relative to the axis of Galactic rotation \citep[cf.][]{Lal:03}. If we assume that it was created by several nearby supernova (SN) explosions in the last $\simeq$$\unit[10]{Myr}$ \citep[e.g.][]{Smi:01}, we are faced with the problem that no young stellar cluster could be found inside its boundaries.

One obvious way out of this dilemma was to scan the solar neighbourhood for moving groups of stars that might have once crossed the 
boundary of the LB and, additionally, provided an adequate number of SNe in order to explain both the present LB extent and the detected soft X-ray emission. For that purpose, \cite{Ber:02} calculated stellar trajectories of the nearby Pleiades subgroup B1 backwards in time. Applying an integer-binned initial mass function (IMF) they inferred the number of SNe in the past, determined their explosion sites from the trajectories, and their explosion times from their main-sequence life times (assuming that all stars in the cluster are born coevally). Similar conclusions were drawn from an independent study by \cite{Mai:01}. Later, \cite{Fuc:06} performed an \emph{ab initio} search scrutinising \emph{all} stars with sufficient kinematical data available within a heliocentric volume of $\unit[400]{pc}$ diameter. Using the HIPPARCOS catalogue and a compilation of radial velocities, they came up with a sample of B stars concentrated in both real and velocity space, which were then identified as members of the Sco-Cen OB association. Tracing back in time this sample, \citeauthor{Fuc:06} found that the stars had entered the present LB volume rather off-centre 10 to $\unit[15]{Myr}$ ago, and that since then, 14 to 20 SN explosions should have occurred, which delivered enough energy to explain the current size of the LB and its ionisation state.

A deeper understanding of the nonlinear evolution and structure of the ISM in general and the LB in particular can only be gained from solving the full-blown set of hydrodynamical (HD) and magnetohydrodynamical (MHD) equations through numerical simulations. While early studies of the global ISM could only consider very few physical processes on small, two-dimensional domains at low numerical resolution \citep{Ban:80,Chi:85,Chi:88,Ros:93}, the advance in computational capabilities not only allowed for the inclusion of additional physics (like magnetic fields) and larger-scale domains \citep{Pas:96,Ros:95}, but eventually also the leap into the third dimension \citep{Low:98,Sto:98,Pad:99}. More recent works can be roughly divided into three classes according to their box size (and the highest resolution associated with it): small-scale \citep{Bal:04,Pio:07}, meso-scale \citep{Avi:04,Avi:05,Sly:05,Gre:08a,Jou:09,Gir:15}, and large-scale \citep{Han:09,Dob:11}. A physically realistic three-dimensional model that covers all dynamical ranges of an entire galaxy is still out of reach, though there are efforts to achieve this to some degree by coupling the adaptive mesh refinement (AMR) technique with zoom-ins to specific regions \cite[see e.g.][and follow-up studies]{Ren:13}.

Taking their modelled ISM as a typical background medium, \cite{Bre:06} further tested the aforementioned LB formation scenario under more realistic conditions. They found that the bubble's extension, as well as its ion column density ratios \citep[see][]{Avi:09,Avi:12}, matched ultraviolet data, as obtained e.g.~with FUSE \citep{Oeg:05,Sav:06}, remarkably well. 

Another interesting way to validate evolution models for the LB is to search for traces of the involved SNe on Earth, manifesting themselves as geological isotope anomalies. \cite{Ell:96} identified several possible target isotopes for such a study, including \element[][60]{Fe} \citep[see also][]{Kor:96}, which is believed to be produced primarily in core-collapse and electron-capture SNe of massive stars, with mass $M\gtrsim \unit[8]{M_\sun}$, via successive neutron capture on preexisting Fe isotopes, both before and during the explosion \citep{Tim:95,Wan:13}. Other suggested sources are asymptotic giant branch \citep[AGB;][]{Bus:03} and super-AGB \citep[SAGB;][]{Lug:12} stars, as well as a rare class of high-density thermonuclear SNe \citep{Woo:97}.

Owing to its comparatively long half-life of $t_{1/2}\simeq\unit[2.6]{Myr}$ \citep{Rug:09,Wal:15}, ejected \element[][60]{Fe} survives long enough not only to travel significant distances (a few hundred pc) through the ISM, but also to be detectable by its $\beta^-$ decay via $\element[][60]{Co}$ and gamma-ray emission at 1173 and $\unit[1333]{keV}$. Measurements with the INTEGRAL satellite \citep{Wan:07} revealed an average photon flux per gamma-ray line of about $\unit[4.4\times 10^{-5}]{cm^{-2}\,s^{-1}\,rad^{-1}}$ for the inner Galaxy region, which might be translated into a steady-state Galactic \element[][60]{Fe} mass of $\unit[1.53]{M_\sun}$ \citep{Die:16}. Very recently, \cite{Bin:16} succeeded in directly detecting \element[][60]{Fe} nuclei in Galactic cosmic rays, showing that \element[][60]{Fe} sources must have been within the distance the high-energy particles can travel for the duration of $t_{1/2}$, which is typically less than about $\unit[1]{kpc}$.

However, the half-life of \element[][60]{Fe} is very short when compared to the age of the solar system ($\simeq$$\unit[4.5]{Gyr}$), implying that the \element[][60]{Fe}, which, as we know from the analysis of primitive meteorites \citep{Tac:03,Mos:05,Biz:07}, was present at about that time, has completely decayed away. This adds to the fact that the isotope's natural terrestrial background, arising mainly from the influx of pre-enriched extraterrestrial material (e.g.~dust and meteorites) onto the Earth, and to a lesser extent from cosmogenic spallative production due to the steady cosmic-ray flux into the Earth's atmosphere, as well as \emph{in situ} production via the penetrating muon and neutron flux, is low \citep{Fie:99}. 

Motivated by this, \cite{Kni:99} used accelerator mass spectrometry to search for an \element[][60]{Fe} excess in a ferromanganese (FeMn) crust from the South Pacific Ocean floor and indeed found one in two layers dating back about $\unit[6]{Myr}$. In a more detailed analysis of another FeMn crust sample (termed 237KD), stemming from the equatorial Pacific, \cite{Kni:04} narrowed down the \element[][60]{Fe} anomaly to the time interval 1.7--$\unit[2.6]{Myr}$ before present \citep[cf.][]{Fei:12}. This definitive signal was corroborated in another study of 237KD conducted by \cite{Fit:08}, who also looked into a different geological reservoir: marine sediments. These grow about a thousand times faster than FeMn crusts and thus have a far better time resolution, albeit the isotopic concentration of each layer is lower. Unfortunately the sediment core analysed by \cite{Fit:08}, which originates from the North Atlantic Ocean and spans the same time interval as the crust 237KD, did not show a compatible \element[][60]{Fe} signal.

In contrast, \cite{Wal:16} found a signal consistent with the expected time from the FeMn crust study in four sediment cores from the Indian Ocean. Remarkably, the signal is broader, rather pointing towards an accumulation of SNe than a single event. Apart from the sediment cores, \citeauthor{Wal:16} also analysed two FeMn crusts (237KD not included) and two FeMn nodules stemming from the Pacific and Atlantic oceans, respectively. Since each of these deep-sea archives contain elevated levels of \element[][60]{Fe} for the time range 1.5--$\unit[3.2]{Myr}$ ago, the authors further concluded that the \element[][60]{Fe} signal must be global. For one of the two FeMn crusts they also detected an additional smaller \element[][60]{Fe} peak 6.5--$\unit[8.7]{Myr}$ ago, whose exact origin is yet rather elusive. Remarkably, a collision of asteroids in the main belt $\unit[8.3]{Myr}$ ago, connected to a boosted, possibly \element[][60]{Fe} enriched, influx of interplanetary dust particles and micrometeorites \citep{Far:06} falls within this particular time range.

Returning briefly to the subject of marine sediments it should be added that these also harbour so-called magnetotactic bacteria, which gather iron from their environment to create nano-sized crystals of magnetite (Fe$_{3}$O$_{4}$). The latter are used by the microbes to orient themselves within Earth's magnetic field, and thus to navigate to their preferred conditions -- a behaviour known as magnetotaxis. \cite{Bis:11} speculated that nearby SN acitivity should have left its mark also in the fossilised remains of such bacteria. In order to prove this hypothesis, they analysed magnetofossils of a sediment core from the eastern equatorial Pacific. The results \citep{Lud:16} indeed show a comparatively weaker \element[][60]{Fe} signal in layers spanning a time range of about $\unit[1]{Myr}$, with the maximum again located at $\simeq$$\unit[2.2]{Myr}$ ago.

In view of all these discoveries it seems little surprising that several of the lunar rocks that were recovered in the course of the Apollo missions also showed an enhanced concentration of \element[][60]{Fe} \citep{Fim:16}. Unfortunately, time-resolved studies are impossible as our cosmic neighbour is regularly hit by a broad range of impactors \citep{Lan:77}, which continuously churn up its surface and mix different layers. Besides this so-called gardening, there is also almost not sedimentation on the Moon. Lunar samples can hence only provide a first hint for the existence of a signature \citep[cf.][]{Fei:13}.

Theoretical attempts to link the FeMn crust measurements to the formation of the LB, but also to a single SN event, were made by \cite{Fei:10} and \cite{Fry:15}, respectively, both on the basis of analytical calculations. \citeauthor{Fei:10} particularly used an SN model developed by \cite{Kah:98}, which does not assume a homogeneous density distribution for the blast wave to expand in, but rather a medium that had already been modified by a previous SN event. More specifically it was assumed that the first SN runs into a homogeneous ISM, whereas all subsequent SNe occur within the spherical region excavated by the first SN, while being exposed to a radial density profile of the form $\rho(R)\propto R^{9/2}$. Following \cite{Fuc:06}, an IMF with one star per mass bin was used, and the most probable explosion sites were calculated. Including isotopic yields from stellar evolution calculations, the measured \element[][60]{Fe} profile could be reproduced with surprisingly high accuracy.

Results of such analytical models should however be treated with some caution, since the similarity solutions applied have a number of serious deficiencies. First, the external pressure is required to be small compared to the pressure in the expanding bubble. This poses no strong limitations for the modelling of a second or subsequent SN event, whose blast wave takes only little time to move through the hot and thin pre-stratified medium. However, the situation is completely different for a single SN, or the very first one in a series, whose blast wave must penetrate through a totally inhomogeneous ISM and thus would soon undergo strong distortions. By the same token, the density distribution of the ambient medium has to be either homogeneous or obey a power law, implying that possible dense obstacles are not taken into account. This includes the LB's outer shell, which was neglected in the preliminarily work by \cite{Fei:10}, but was later implemented as a \cite{Wea:77} wind bubble for the improved analytical study contained in \citet[hereafter BFS+]{Bre:16}. Second, the external medium is taken to be constant over time. In reality, however, the medium changes after each SN. \citeauthor{Fei:10} (and also \citetalias{Bre:16}) at least switched from a homogeneous to a power-law density distribution after the first SN went off. Nevertheless, they assumed that all subsequent explosions occur in the same environment. Last but not least, turbulent mixing and mass loading, as present in real SBs, are not taken into account in those models.

A definitive answer to the question whether the observed \element[][60]{Fe} excess is explainable in the context of the LB formation can therefore only be given through three-dimensional high-resolution numerical simulations. This paper details on our first step working towards that goal \citep[see also][]{Schu:15}. For this purpose we built upon the numerical simulations of \citetalias{Bre:16}, presenting an additional homogeneous model (see Table \ref{tab:bgmprops}, model A). As it will be shown this sets a lower limit for possible homogeneous media. We particularly used the masses of the same 16 SN progenitor stars, together with the explosion times, total ejected masses, and \element[][60]{Fe} mass fractions derived from there. The applied SN explosion sites, on the other hand, result from the most probable stellar trajectories (see also \citetalias{Bre:16}). 
Like \cite{Bre:06}, we studied the formation and evolution of the LB not in isolation, but in concert with the neighbouring SB Loop~I. For the sake of conciseness, we currently neglected the complex multiphase nature of the (local) ISM and considered instead only homogeneous background media, including the one already used in \citetalias{Bre:16}. Deeper conclusions can be drawn now, however, as we not only traced the global  \element[][60]{Fe} dynamics in our solar neighborhood but also the contributions of the individual SNe. This was achieved by tagging each of them with their own passive scalar (or tracer) -- a quantity that behaves like a drop of dye when dispersed in a liquid. The influence of the inhomogeneous SN-driven environment, already teased in \citetalias{Bre:16}, will be extensively analysed in the forthcoming Paper II, which will also contain a discussion of the properties of turbulence in such a self-consistently evolved medium.

The article is organised as follows. Section \ref{ch:model} describes the model setup, in Sect.~\ref{ch:results} our results with respect to SB evolution and comparison to the FeMn crust measurements are discussed, and we close with a general discussion and conclusions in Sect.~\ref{ch:concl}.
%

\section{Model description}
\label{ch:model}
Our HD simulations were carried out with the tree-based AMR code \textsc{Ramses} \citep{Tey:02}, which allows for solving the discretised Euler equations in their conservative form by means of a second-order unsplit Godunov method for perfect gases. In particular, we employed the MUSCL-Hancock scheme  \citep{Lee:79} together with the Local Lax-Friedrichs (or Rusanov) approximate Riemann solver \citep{Lax:54,Rus:61} and the MinMod slope limiter \citep{Roe:86}. The computational domain is cubic with $\unit[3]{kpc}$ side length and covered by a base cartesian grid of $64^3$ uniform cells. In regions of steep density and pressure gradients, the grid was recursively refined up to six additional levels, resulting into a peak spatial resolution of $\unit[0.7]{pc}$. This high resolution was also kept at the coordinate origin, which marked the location of Earth (or Sun; for a distinction one would require several additional levels of grid refinement, and thus much more computing time). The grid was initially filled with a uniform and static medium. The adopted values for the gas temperature $T$, particle density $n$, and metallicity (in solar units) $Z/Z_\sun$ in our two best-fit models are given in Table \ref{tab:bgmprops}. They actually mimic two specific `phases' of the classical three-phase ISM model by \cite{Kee:77}, namely the warm ionised medium (model A) and the warm neutral medium (model B). The SB-generating SNe were initialised as blast waves right at the start of their Sedov-Taylor phase, each liberating a canonical explosion energy of $E_\mathrm{SN}=\unit[10^{51}]{erg}$. The self-gravitating gas was exposed to an isotropic radiation field (that includes absorption through the Galactic plane as well as the cosmic microwave background), but was also allowed to loose energy via various interstellar cooling processes. This energy gain and loss was modelled under the assumption of collisional ionisation equilibrium (CIE) using the temperature-dependent net cooling functions generated for several gas densities with the spectral synthesis code \textsc{Cloudy} \citep{Ferl:98}. The  boundaries of our computational domain were formally periodic, but were actually never reached by expanding shells during the time frame considered. As a further `boundary condition', we set up a neighbouring SB, intended to represent Loop~I. This is required since ROSAT observations have shown that soft X-rays are absorbed by a nearby neutral `wall' enclosed by an even denser ring-like structure, which is most likely the result of an interaction between the LB and the Loop~I SB \citep{Egg:95}.

\begin{table}
\caption{Background media properties.}
\label{tab:bgmprops}
\centering
\begin{tabular}{cccc}
\hline\hline
Model & $T$ & $n$ & $Z/Z_\sun$\\
          & $(K)$ & $(cm^{-3})$ & \\
\hline
A & 10\,000 & 0.1 & 1\\
B & 6800 & 0.3 & 1\\
\hline
\end{tabular}
\end{table}

\onltab{
\begin{table*}
\caption{Loop~I SB model parameters.} 
\label{tab:loopipars}
\centering
\begin{tabular}{ccccccccl}
\hline\hline
$t_{\mathrm{exp}}$ & $x$  & $y$  & $z$  & $d$ & $M_*$               & $M_{\mathrm{ej}}$ & ${Z}$          & Subgroup \\
(Myr)                       & (pc)  & (pc)  & (pc) & (pc) & ($M_{\sun}$) & ($M_{\sun}$)           & ($10^{-6}$) &       \\
\hline
$-11.9$ & $309$ & $-229$ & $ 17$ &  $385$ & $18.42$ & $13.19$ & $4.171$ & Vel OB2\\
$-11.2$ & $275$ & $-237$ & $ 17$ &  $363$ & $17.14$ & $12.40$ & $3.947$ & Vel OB2\\
$-10.5$ & $245$ & $-255$ & $ 17$ &  $354$ & $16.10$ & $11.75$ & $3.786$ & Vel OB2\\
$- 9.8$  & $226$ & $-312$ & $  8$ &  $385$ & $15.18$ & $11.19$ & $3.660$ & Vel OB2\\
$- 9.2$  & $202$ & $-334$ & $-14$ &  $391$ & $14.37$ & $10.69$ & $3.560$ & Vel OB2\\
$- 8.5$  & $194$ & $-342$ & $ 10$ &  $393$ & $13.64$ & $10.24$ & $3.479$ & Vel OB2\\
$- 7.8$  & $196$ & $-313$ & $-45$ &  $372$ & $12.98$ & $ 9.83$ & $3.414$ & Vel OB2\\
$- 7.2$  & $134$ & $-343$ & $12$ &  $368$ & $12.39$ & $ 9.47$ & $3.362$ & Vel OB2\\
$- 6.6$  & $137$ & $-348$ & $ 72$ &  $381$ & $11.86$ & $ 9.14$ & $3.320$ & Tr 10\\
$- 5.9$  & $ 80$  & $-222$ & $ -4$ &  $236$ & $11.36$ & $ 8.84$ & $3.285$ & Vel OB2\\
$- 5.3$  & $ 97$  & $-238$ & $-25$ &  $258$ & $10.91$ & $ 8.56$ & $3.257$ & Vel OB2\\
$- 4.6$  & $ 84$  & $-279$ & $-18$ &  $292$ & $10.50$ & $ 8.30$ & $3.235$ & Vel OB2\\
$- 4.0$  & $ 73$  & $-327$ & $ 42$ &  $337$ & $10.12$ & $ 8.07$ & $3.218$ & Tr 10\\
$- 3.4$  & $ 67$  & $-325$ & $ 36$ &  $334$ & $ 9.76$ & $ 7.85$ & $3.204$ & Tr 10\\
$- 2.8$  & $ 60$  & $-310$ & $-26$ &  $317$ & $ 9.43$ & $ 7.64$ & $3.194$ & Vel OB2\\
$- 2.1$  & $   9$  & $-282$ & $-27$ &  $283$ & $ 9.13$ & $ 7.46$ & $3.186$ & Vel OB2\\
$- 1.5$  & $-20$  & $-262$ & $-19$ &  $264$ & $ 8.84$ & $ 7.28$ & $3.181$ & Vel OB2\\
$- 0.9$  & $-11$  & $-273$ & $-48$ &  $277$ & $ 8.57$ & $ 7.11$ & $3.178$ & Vel OB2\\
$- 0.3$  & $-15$  & $-279$ & $  7$ &  $280$ & $ 8.32$ & $ 6.96$ & $3.176$ & Tr 10\\
\hline
\end{tabular}
\tablefoot{
The first column contains the Loop~I SN explosion times $t_{\mathrm{exp}}$ (counted backwards from today). In columns 2--5, we give the explosion centres in cartesian coordinates $(x,y,z)$ with the Sun at the origin, and the resulting distances of the SNe from the Sun $d$. Columns 6--8 give the masses of each SN progenitor star $M_*$, the masses ejected in each SN explosion $M_{\mathrm{ej}}$, and the corresponding $^{60}$Fe mass fractions ${Z}$. Column 9 lists the parent subgroup of each progenitor star.
}
\end{table*}
}
%

\subsection{Setting up the Loop~I superbubble}
\label{ch:loopI}
For identifying possible SN progenitor stars of Loop~I, we performed a search for B stars analogous to the one of the LB \citep{Fuc:06}, but for a spherical search volume of twice the size. Since less kinematical data is available for stars at larger distances, it becomes increasingly difficult to find bright stars. It turns out that Tr 10 and the association Vel OB2 have recently passed through the present volume of Loop~I rather off-centre. Like for the LB this is not a problem \emph{per se}, as the extension and morphology of evolved SBs are primarily determined by the ambient density and pressure distribution.

Given a sample of 80 stars that might have entered this volume about 12.3 Myr ago (main sequence lifetime of a still existing $\unit[8.2]{M_{\sun}}$ star), we first needed to estimate how many have already exploded. For that purpose, we fitted an IMF that is typical for young massive stars with initial masses $M$,    
\begin{equation}
\frac{\mathrm{d}\mathcal{N}}{\mathrm{d}M}=\left. \frac{\mathrm{d}\mathcal{N}}{\mathrm{d}M}\right\vert_{0}M^{\Gamma-1}\,,
\label{eq:massey}
\end{equation}
where $\Gamma=-1.1\pm 0.1$ \citep{Mas:95,Ber:02,Fuc:06}. With the lower and upper limit of the relevant mass range being defined by A0 and B0 stars, respectively (i.e.~$M_{\mathrm{l}}=\unit[2.6]{M_{\sun}}$ and $M_{\mathrm{u}}=\unit[8.2]{M_{\sun}}$), we obtained the normalisation constant via integration of Eq.~(\ref{eq:massey});
\begin{equation}
\begin{split}
\left. \frac{\mathrm{d}\mathcal{N}}{\mathrm{d}M}\right\vert_{0}&=\mathcal{N}\left[\int_{2.6}^{8.2}M^{-2.1}\,\mathrm{d}M\right]^{-1}\\
& = 80\left[\frac{2.6^{-1.1}-8.2^{-1.1}}{1.1}\right]^{-1}= 351\,.
\end{split}
\end{equation}

The lifespan of the first and most massive star that exploded in Loop~I might be expressed as
\begin{equation}
\tau_{M_{t'}}=\tau_\mathrm{u}-t'\,,
\label{eq:taumdeltatau}
\end{equation}
where $\tau_\mathrm{u}$ is the lifetime of a star of mass $M_\mathrm{u}$ and $t'$ is the time at which the association entered the present volume of Loop~I. When quantifying the main sequence lifetime of stars in the mass range $2\le M\le \unit[67]{M_\sun}$ by a fit to the isochrone data of \cite{Sch:92},
\begin{equation}
\tau=\tau_0 M^{-\beta}\,,
\label{eq:mslifetime}
\end{equation}
where $\tau_0=\unit[1.6\times 10^8]{yr}$ and $\beta=0.932$ \citep{Fuc:06}, Eq.~(\ref{eq:taumdeltatau}) translates into
\begin{equation}
M_{t'}=\left(M_\mathrm{u}^{-\beta}-\frac{t'}{\tau_{0}} \right)^{-1/\beta}\,.
\end{equation}
As $t'=\unit[12.3]{Myr}$, the mass of the most massive star that has already exploded is $M_{t'}=\unit[19.2]{M_{\sun}}$. One therefore finds for the number of missing stars
\begin{equation}
\begin{split}
\mathcal{N}_{\mathrm{SN}}&=\int_{8.2}^{19.2}\left. \frac{\mathrm{d}\mathcal{N}}{\mathrm{d}M}\right\vert_{0}M^{-2.1}\,\mathrm{d}M\\
&=351\left[\frac{8.2^{-1.1}-19.2^{-1.1}}{1.1}\right]=19\,.
\label{eq:NSN}
\end{split}
\end{equation}

Obviously the masses of these SN progenitors should be based on the IMF introduced above. In order to obtain the statistically most probable distribution \citep{Mai:05}, we performed the mass binning by assigning exactly one star to each mass interval of the IMF, where the actual stellar masses are taken to be the \emph{average} mass in each bin. Analogously, \cite{Fei:10} (and also \citetalias{Bre:16}) derived the initial masses of the LB stars, which we also employed in this work. Using Eq.~(\ref{eq:massey}), we obtained for the mass of a particular star lying in the mass bin between $M_1$ and $M_2$ ($>$$M_1$),
\begin{equation}
\begin{split}
M_{*}&=\frac{1}{2}\left(M_{1}+M_{2}\right)\\
&=\frac{1}{2}\left\{M_{1}+\left[M_{1}^{-1.1}-1.1\left(\left. \frac{\mathrm{d}\mathcal{N}}{\mathrm{d}M}\right\vert_{0}\right)^{-1}\right]^{-1/1.1}\right\}\,.
\label{eq:binmass}
\end{split}
\end{equation}

These initial masses served as a basis for determining the total ejected masses \citep[using stellar evolution data from][]{Woo:07} as well as the \element[][60]{Fe} mass fractions (see Sect.~\ref{ch:fe60_calc}). The explosion times, on the other hand, are defined by $t_{\mathrm{exp}}=\tau-\tau_{\mathrm{c}}$, where $\tau$ is as given in Eq.~(\ref{eq:mslifetime}) and $\tau_{\mathrm{c}}$ is the age of the cluster, which should be comparable to the lifespan of a $\unit[8.2]{M_{\sun}}$ star, i.e.~$\tau_{\mathrm{c}}=\unit[23]{Myr}$. For arranging the stellar explosion centres in space, we calculated the trajectories of all known B stars of Tr 10 and Vel OB2. Since not all of the member stars pass through the Loop~I volume we selected for the 19 SN progenitors only the closest to the centre. A full list of the Loop~I input parameters is given in Table \ref{tab:loopipars} (available electronically only).
%

\subsection{Calculating the amount of SN-released \element[][60]{Fe} that arrives on Earth}
\label{ch:fe60_calc}
According to stellar evolution models \citep[e.g.][]{Lim:06,Woo:07}, there is a close link between the \element[][60]{Fe} content of a star and its mass. For the considered stellar mass range of $\simeq$10--$\unit[30]{M_\sun}$, the calculated \element[][60]{Fe} yields however exhibit a large scatter between $\simeq$$10^{-6}$ and $\unit[10^{-4}]{M_\sun}$. This is due to the sensitivity of nucleosynthesis to a variety of factors, including cross-sections, mass loss, and rotation. For our model, we neglected these subtleties and instead used the exponential fit to recent stellar evolution data provided in \citetalias{Bre:16}.

Owing to its low concentration, \element[][60]{Fe} should have no dynamical influence on the (turbulent) flow. We therefore treated it as a passive scalar, obeying an advection-diffusion equation of the form \citep[see e.g.][]{Dav:04}
\begin{equation}
\frac{\partial Z}{\partial t}+(\vec{u}\cdot \boldsymbol{\nabla})Z=\alpha \nabla^2 Z\,,
\label{eq:Zadvdiff}
\end{equation}
where $Z$ is the \element[][60]{Fe} mass fraction, $\vec{u}$ the fluid velocity, and $\alpha$ the diffusivity of the contaminant (which is assumed here to be isotropic). Since for the case of the ISM the Peclet number, $\mathrm{Pe}=UL/\alpha$, is large ($U$ and $L$ denote characteristic velocity and length scales), diffusive effects are restricted to the very small, so-called microscale of turbulence. At larger scales, the right-hand side of Eq.~(\ref{eq:Zadvdiff}) can hence be neglected. The scalar $Z$ then acts like a marker that tags the chemically enriched fluid parcels. However, there is still a need for some diffusive process ultimately operating at small scales, even though its exact value is quite insignificant for the turbulent mixing at the larger scales, as demonstrated by \cite{Avi:02} in ISM simulations with different resolutions. Like them we did not implement any physical diffusion term, but rather left the mixing to numerical diffusion, which is generally faster and operates on a larger scale than its physical counterpart. As a consequence, the timescale of mixing in our simulations rather poses a lower limit to the mixing time resulting from physical diffusion.

When generalising the analysis to the spatiotemporal evolution of $m$ passive scalars with concentrations $Z_i$ $(i = 1,...,m)$, the time-dependent Euler equations in conservation-law form read as \citep[see e.g.][]{Tor:09}
\begin{equation}
\frac{\partial \vec{U}}{\partial t}+\frac{\partial \vec{F}(\vec{U})}{\partial x}+\frac{\partial \vec{G}(\vec{U})}{\partial y}+\frac{\partial \vec{H}(\vec{U})}{\partial z}=\vec{S}(\vec{U})\,,
\end{equation}
with the state vector
\begin{equation}
\vec{U}= \left[\rho,\rho u,\rho v,\rho w, E,\rho Z_1,\dots,\rho Z_m\right]^\mathrm{T}\,,
\end{equation}
the vectors of fluxes in the corresponding coordinate directions
\begin{align}
\vec{F}&=\left[\rho u,\rho u^2+P,\rho uv,\rho uw, u(E+P),\rho u Z_1,\dots,\rho u Z_m\right]^\mathrm{T}\,,\\
\vec{G}&=\left[\rho v,\rho uv,\rho v^2+P,\rho vw, v(E+P),\rho v Z_1,\dots,\rho v Z_m\right]^\mathrm{T}\,,\\
\vec{H}&=\left[\rho w,\rho uw,\rho vw,\rho w^2+P, w(E+P),\rho w Z_1,\dots,\rho w Z_m\right]^\mathrm{T}\,,
\end{align}
and $\vec{S}(\vec{U})$ denoting an appropriate source term vector. Here, $\rho$ is the fluid density, $u$, $v$, $w$ the components of the velocity vector $\vec{u}$, $E$ the total energy, and $P$ the (thermal) pressure. Imprinted in this evolution is the radioactive decay of \element[][60]{Fe}, for which we also accounted in our simulations. The quantitative comparison between our model results and the FeMn crust measurements was then accomplished via the following steps.

First, we determined the local interstellar fluence, $\mathcal{F}$, which is the flux of the isotope at the Earth's position integrated over time. Since this position coincides with the point of contact of eight, highest resolved grid cells, with the HD variables being always defined at the cell centres, we had to average over the \element[][60]{Fe} mass fluxes, $\rho |\vec{u}|Z$, in those cells. Hence,
\begin{equation}
\mathcal{F}=\frac{(\rho |\vec{u}|Z)_{\mathrm{VA}}}{A m_{\mathrm{u}}}\Delta t\,,
\label{eq:feflu}
\end{equation}
where $(\dots)_{\mathrm{VA}}$ denotes the volume average, $A=60$ the mass number of \element[][60]{Fe}, $m_{\mathrm{u}}\simeq \unit[1.66\times 10^{-24}]{g}$ the atomic mass unit, and $\Delta t$ the last simulation time step. 

For the remaining steps we took the fall-out on Earth to be isotropic, which was demonstrated by \cite{Fry:16} to be a suitable assumption. The surface density of \element[][60]{Fe} atoms deposited at time $t$ before present in the FeMn crust can then be expressed as\footnote{Sometimes in the literature the quantity $\Sigma$ is referred to as the `observed fluence' \citep[cf.][]{Fry:15}.} \citep[cf.][]{Fie:99}
\begin{equation}
\Sigma =\frac{fU}{4}\mathcal{F} \exp(-\lambda t)\,,
\label{eq:fesurfden}
\end{equation}
where the factor $\nicefrac{1}{4}$ stems from relating Earth's cross-section ($R_{\oplus}^{2}\pi$) to its surface area ($4 R_{\oplus}^{2}\pi$), the exponential term describes the decay of the radionuclide during its final disposal in the FeMn crust ($\lambda\equiv\ln(2)/t_{1/2}$), and the product $fU$ represents the \element[][60]{Fe} survival fraction. 

Specifically, $f$ denotes the fraction of \element[][60]{Fe} atoms that reaches the Earth's orbit after overcoming a variety of filtering processes, while being condensed into dust grains. After estimating the dust condensation of \element[][60]{Fe} at departure from source, as well as its destruction through interstellar passage, heliosphere, and solar radiation pressure, \cite{Fry:15} concluded that $f\simeq 0.01$. 

The other, so-called uptake factor, $U$, accounts for the fact that only a certain fraction of \element[][60]{Fe} spread over Earth's surface actually finds its way into the FeMn crust. The underlying chemical selection process is only poorly understood, implying that the value of $U$ is, as of $f$, very uncertain. \cite{Kni:04} quantified $U$ using the relative concentrations of iron and manganese in water and the FeMn crust, and the uptake of manganese (4\,\%), while implicitly assuming that $f=1$. They found $fU\equiv U\simeq 0.006$. More recent studies, however, suggest that $U=0.5$--1 \citep{Bis:11,Fei:12}. Hence, by taking $U=0.6$ and $f$ as derived by \cite{Fry:15}, one retains the numerical value claimed by \cite{Kni:04}, but still adds some (very simplistic) sub-grid dust physics to the problem.

We finally obtained the number density of \element[][60]{Fe} for each crust layer by summing $\Sigma$ in the time bins lying within the corresponding time intervals and then dividing by the thickness of the layer. Measurements demand to relate these densities to the one of stable iron, i.e.~$n_{^{60}\mathrm{Fe}}/n_{\mathrm{Fe}}$, or, in short, \element[][60]{Fe}/Fe. The required normalisation constant follows from \citep[cf.][]{Fei:14}
\begin{equation}
n_\mathrm{Fe}=\frac{w_\mathrm{Fe} \rho N_\mathrm{A}}{A}\,,
\end{equation}
where $w_\mathrm{Fe}\simeq 0.1527$ is the weight fraction of iron in the FeMn crust sample whose mass density $\rho$ is $\unit[1.5]{g\,cm^{-3}}$, $N_\mathrm{A}\simeq \unit[6.022\times 10^{23}]{mol^{-1}}$ is the Avogadro constant, and $A\simeq\unit[55.845]{g\,mol^{-1}}$ is the molar mass for iron. We thus have $n_{\mathrm{Fe}}\simeq\unit[2.470\times 10^{21}]{cm^{-3}}$.

\subsection{Overview of input data}
To conclude this section we give a full list of the input data and recapitulate how it is obtained:
\begin{itemize}
\item Sample of cluster stars: selected according to compactness in real and velocity space from astrometric catalogue data (such as HIPPARCOS) for a heliocentric sphere with diameter $D\sim \unit[400]{pc}$ (LB) and $\unit[800]{pc}$ (Loop~I)
\item Cluster age ($\tau_\text{c}$): derived by comparison of the cluster turn-off point with isochrones of \cite{Sch:92}
\item Number and initial masses of SN progenitors ($\mathcal{N}_\text{SN}$, $M_*$): estimated via IMF fitting of sample stars \citep{Mas:95}, assuming one star per mass bin with the bin's average mass for the perished stars [Eqs.~(\ref{eq:NSN}) and (\ref{eq:binmass})]
\item Main sequence lifetimes of SN progenitors ($\tau$): calculated as a function of mass only via Eq.~(\ref{eq:mslifetime})
\item SN explosion times ($t_\text{exp}$): calculated from subtracting $\tau_\text{c}$ from $\tau$, thus assuming coeval star formation in the cluster
\item SN explosion centres ($x,y,z$): derived from the center-of-mass stellar trajectories, as determined via the epicyclic equations of motion \citep{Lin:59,Wie:82}; for the LB, the most probable ones are calculated for a Gaussian error distribution in the HIPPARCOS positions and proper motions
\item Total ejecta mass and \element[][60]{Fe} mass fractions ($M_\text{ej}$, $Z$): taken, corresponding to $M_*$, from stellar evolution models \citep{Woo:95,Rau:02,Lim:06,Woo:07}
\item \element[][60]{Fe} survival fraction ($fU$): quantified by combining estimates from a dust survival model \citep[][and references therein]{Fry:15} and measurements in terrestrial archives \citep{Kni:04,Bis:11,Fei:12}; the dynamics of dust and gas is assumed to be perfectly coupled, where the latter is traced via passive scalars
\end{itemize}
%

\section{Results}
\label{ch:results}

\subsection{Evolution and properties of the Local and Loop~I superbubbles}
\label{ch:evo_AB}

   \begin{figure*}
   \resizebox{\hsize}{!}
            {\includegraphics{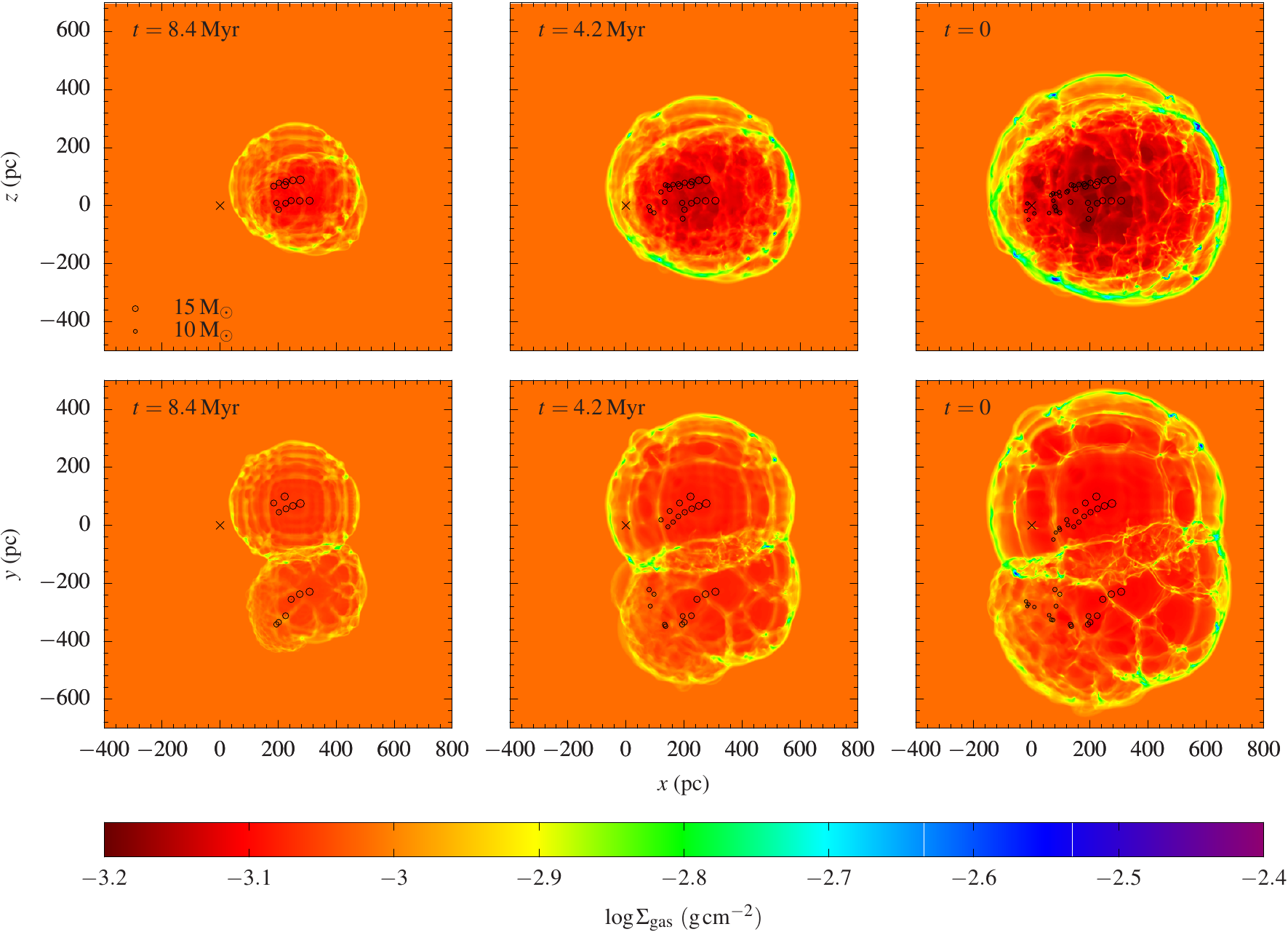}}
      \caption{Colour-coded maps of the logarithmic total gas column density in model A for three different times before present. Integrations along the $y$- and $z$-axis are shown in the upper and lower panels, respectively. Earth's projected position is marked by a cross. Circles indicate the projected centres of SN explosions that occurred in the time frame before the snapshot was taken. The sizes of the circles correspond to the initial masses of the progenitor stars (see legend in the upper left panel).}
         \label{im:colmaps_A}
   \end{figure*}

   \begin{figure*}
   \resizebox{\hsize}{!}
            {\includegraphics{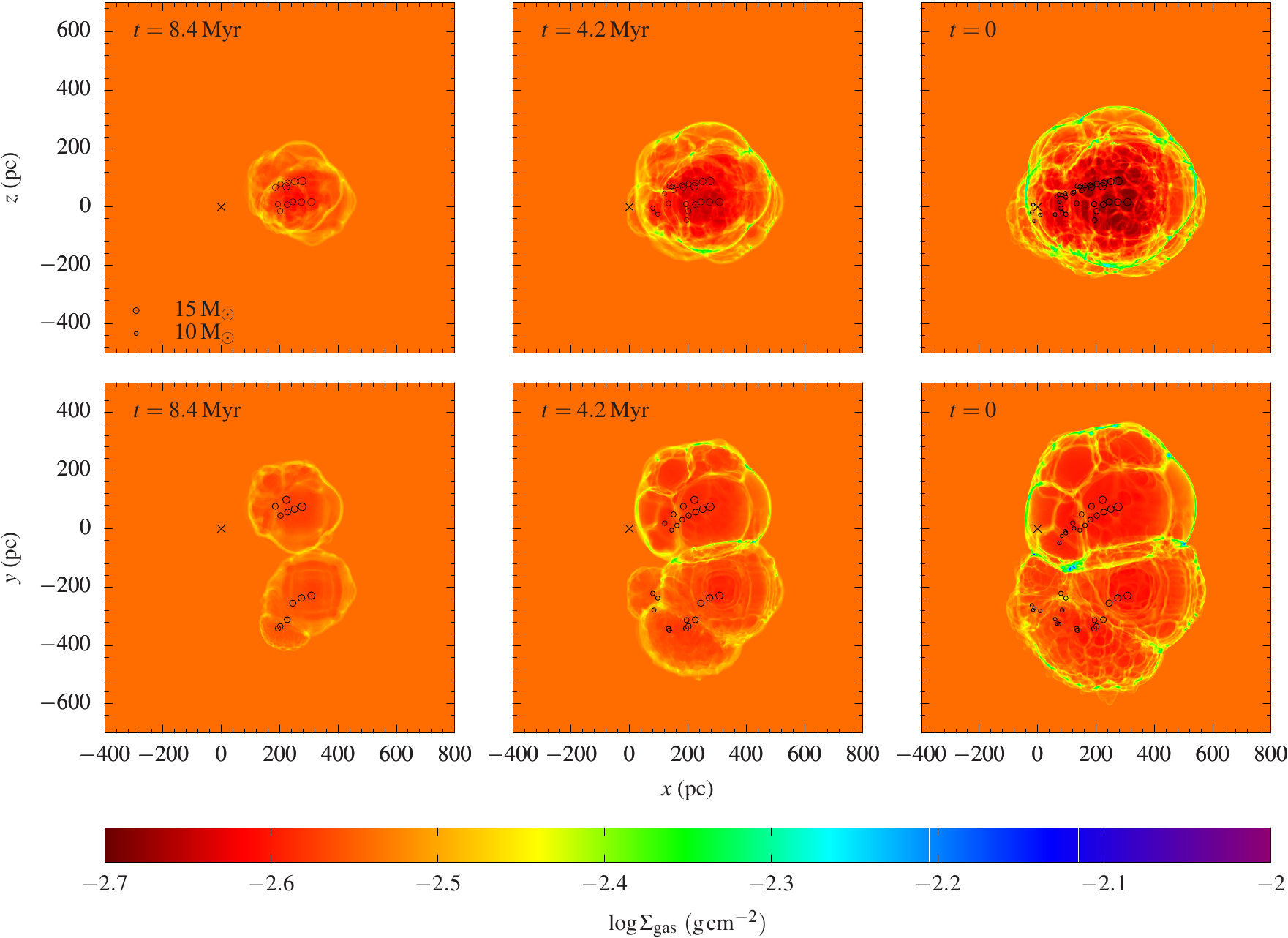}}
      \caption{As for Fig.~\ref{im:colmaps_A}, but for model B.}
         \label{im:colmaps_B}
   \end{figure*}

   \begin{figure*}
   \resizebox{\hsize}{!}
            {\includegraphics{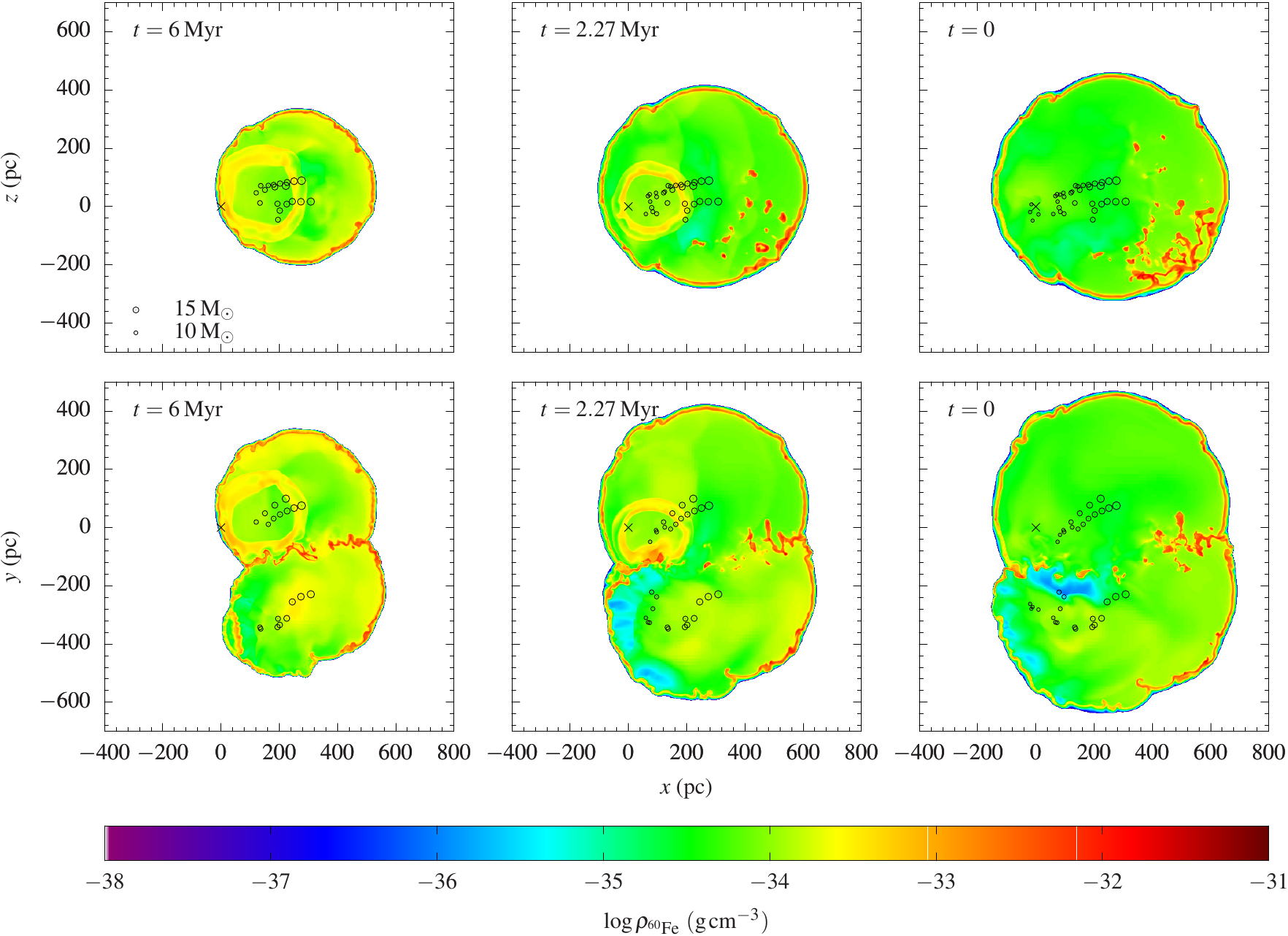}}
      \caption{Colour-coded maps of the logarithmic  \element[][60]{Fe} mass density in model A at three different times before present. Vertical ($y = 0$) and horizontal ($z = 0$) cuts through the computational box are shown in the upper and lower panels, respectively. Symbols are as in Figs.~\ref{im:colmaps_A} and \ref{im:colmaps_B}.}
         \label{im:fedmaps_A}
   \end{figure*}

      \begin{figure*}
   \resizebox{\hsize}{!}
            {\includegraphics{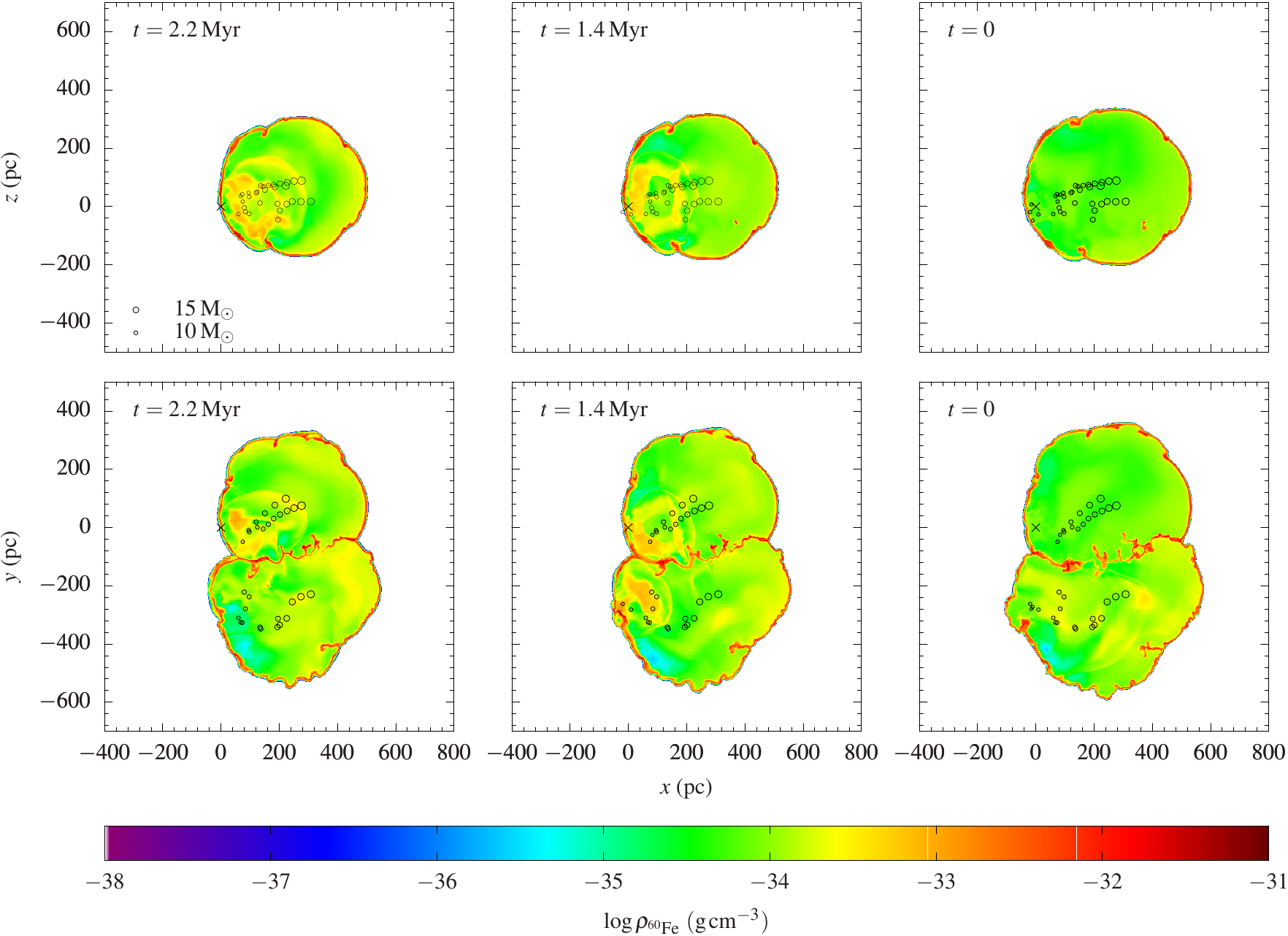}}
      \caption{As for Fig.~\ref{im:fedmaps_A}, but for model B.
              }
         \label{im:fedmaps_B}
   \end{figure*}

The SN explosions generate a coherent bubble structure whose evolution is followed over $\unit[12.6]{Myr}$ (age of the LB) until the present time. In Figs.~\ref{im:colmaps_A} and \ref{im:colmaps_B} we present snapshots of the total gas column density for the models A and B, respectively. Note that in these, as well as in all remaining maps, the $x$, $y$, and $z$-axes point towards the Galactic centre, into the direction of the Galactic rotation, and towards the Galactic north pole, respectively.

After their almost coeval formation, the two SBs evolve at first completely independently from each other. The shocked ambient medium cools very quickly and gets compressed into thin shells. These supershells are subject to thermal and fluid dynamical instabilities, which give rise to the formation of cold and dense clumps. In particular, during the `breaks' between two consecutive SN explosions, when the supershell is decelerating, Rayleigh-Taylor instabilities develop at the contact discontinuity that separates the shocked SN ejecta from the cooler and denser swept-up ambient medium. The filaments of dense gas produced thereby grow into the cavity where they can mix with the hot bubble gas. In addition, an overstability sets in that arises from a mismatch between thermal and ram pressure at the shell surface. This `Vishniac instability' \citep[see][and references therein]{ryu:91} can cause growing, oscillating ripples on that surface. Although such structures are particularly prominent for extinct SBs, whose shell motion has stalled due to the lack of energy sources \citep[i.e.~stellar winds and SNe; cf.][]{Kra:13}, they may already be visible at the late stages of the simulations presented in this work; see e.g.~the spike protruding from the shell of Loop~I at $(x,y)=\unit[(240,-600)]{pc}$ in the lower right panel of Fig.~\ref{im:colmaps_B}. It is also interesting to note a criss-cross network of shocks as a result of sequential SN explosions. This may lead to stronger coalescing shock waves and thus to local pressure enhancements, which may also be responsible for local deviations from sphericity of the LB and Loop~I despite the homogeneous ambient medium.

Certainly an effect of the latter is that the supershells retain their spherical shape on large scales. After about $\unit[3]{Myr}$ ($t\simeq\unit[9.6]{Myr}$ before present; model A) and $\unit[4.6]{Myr}$ ($t\simeq\unit[8]{Myr}$ before present; model B) the shells collide, giving rise to an even denser and cooler interaction layer. Also this interface becomes eventually Rayleigh-Taylor-unstable, since the pressure in the two SB volumes is not equal. This leads to the formation of cold and dense cloudlets that travel into the direction of lower pressure (i.e.~the LB interior) as soon as the instability becomes fully non-linear \citep[cf.][]{Bre:00}. In model A, the interaction shell breaks up after an evolution time of about $\unit[6.5]{Myr}$ ($t\simeq\unit[6.1]{Myr}$ before present), allowing for immediate exchange of hot gas between the SBs. In model B, on the other hand, no break-up occurs during the whole evolution time considered. The final size of the LB is about $800$ by $\unit[600]{pc}$ (model A) and $580$ by $\unit[480]{pc}$ (model B) in the $(x,y)$-plane, and about $\unit[760]{pc}$ (model A) and $\unit[540]{pc}$ (model B) perpendicular to it. The atomic hydrogen number density and temperature of the LB interior, particularly in the solar neighbourhood, is about $10^{-4.2}$--$\unit[10^{-3.9}]{cm^{-3}}$ and $10^{6.9}$--$\unit[10^{7.1}]{K}$ in model A, and about $10^{-4.2}$--$\unit[10^{-3}]{cm^{-3}}$ and $10^{5.8}$--$\unit[10^{7}]{K}$ in model B. If the ambient medium is homogeneous, it is generally the first SN that determines the final appearance of the SB most. This is because within a few sound-crossing times the bubble has a homogeneous pressure acting on the shell of the first explosion. The centre of the final SB should consequently lie very close to the explosion centre of the earliest SN. That this is indeed the case, also in our models, can be seen best in the bottom panels of Figs.~\ref{im:colmaps_A} and \ref{im:colmaps_B} (the largest black circle marks the locations of the earliest explosion). Furthermore, since we used the same SN data for both models, the relative arrangement of the two SBs does not change either.

It should be noted that the calculated final $x$- and $y$-extensions only poorly match the sizes derived from observations. Also the elongation in $z$-direction is highly underestimated. This is because the background medium is assumed to be homogeneous, which is certainly an oversimplification of the actual conditions in the solar neighbourhood. As seen from the extended, three-dimensional differential opacity maps of the local ISM by \cite{Lal:14}, the LB is bound today by a series of dense clouds located in the first and fourth quadrants, as well as in the anti-centre region of the Galactic plane. These clouds, which also possess non-negligible vertical extensions, are known to be part of Gould's belt. Moreover, there is a 500--$\unit[1000]{pc}$ wide cavity in the third quadrant, as well as a small, elongated cavity in the opposite direction. The larger cavity is centred below the Galactic plane and was identified as the counterpart of the so-called GSH 238+00+09 supershell, which is visible in radio wavelengths \citep[see][and references therein]{Pus:14}. Assuming that these structures were present in a similar fashion as observed today throughout the entire evolution time, the LB's outer shell would probably have experienced early on a greater `resistance' in the second, fourth, and (partly) the first quadrant, and would thus not have grown to such a great extent in those directions. This should particularly hold for model A, in which the SBs expand more rapidly. The evolution in the third quadrant is more difficult to predict since in our models, this region is partly occupied by Loop~I. The presence of the dense interstellar cloud at a distance of $\simeq$$\unit[200]{pc}$ and longitude $240^{\circ}$ might nevertheless restrict the LB growth in a similar fashion as the developing interaction layer in our simulations. The combined effect of all these structures might also `shift' the solar system closer to the centre of the then smaller LB. Further knowledge can only be gained from additional simulations whose initial conditions are tailored to match these recent observations. These will be the subject of a forthcoming paper.

Current results however suggest that the exact size and shape of the LB does not play a great role for modelling the \element[][60]{Fe} amount that arrives on Earth, except for the case in which \element[][60]{Fe} is carried to the solar system by a reflected shock (see Sec.~\ref{ch:compa}). Generally it is necessary that the solar system lies within the LB so that it can be reached by the expanding SN remnants. For this, the gas-dynamical properties of the medium between the SN explosion centres and the locus of the Earth are crucial, as they determine the arrival time of the LB's outer shell and hence set the date for the earliest possible \element[][60]{Fe} signal in the terrestrial record.

We further note that the relative locations of the Local and the Loop~I SB are not the same as in the joint-evolution model by \cite{Bre:06}. This is because they used different stellar trajectories for the LB progenitors, and, more importantly, a different formation scenario for Loop~I, based on other moving group stars from the Sco-Cen cluster for which they assumed a single explosion centre that is somewhat arbitrarily shifted $\unit[200]{pc}$ relative to the centre of the LB in positive $x$-direction. Therefore, and contrary to our model, their LB is generated by 19 SNe, and their Loop~I SB by 39 SNe. Apart from that there is some discussion in the literature whether the large circular feature in the radio continuum sky, known as the North Polar Spur, which is associated with Loop~I, is part of the local ISM at all or rather belongs to the Galactic centre region \citep[see][and references therein]{Pus:14}. The presence of nearby young stars in the direction of the Galactic centre, however, strongly argues for SN explosions that generated the Loop~I SB.

Using a passive scalar as discussed in Sect.~\ref{ch:fe60_calc}, we followed the spatiotemporal evolution of the SN-released \element[][60]{Fe}. Corresponding simulation snapshots are shown in Figs.~\ref{im:fedmaps_A} and \ref{im:fedmaps_B} for models A and B, respectively. From these maps it is seen that inhomogeneities in the \element[][60]{Fe} distribution, arising from recent SNe, are smoothed out over time. Responsible for this mixing process are turbulent shear flows resulting from SN blast waves that run into the surrounding supershell and thus generate asymmetric reflected shock waves. The zones of underpressure left behind by these reflected discontinuities manifest themselves as the blue `lakes' particularly visible in the lower panels of Figs.~\ref{im:fedmaps_A} and \ref{im:fedmaps_B}. With a characteristic large-eddy size of $\ell\simeq\unit[100]{pc}$ and an average sound speed after an SN of the order of $a\simeq\unit[10^2]{km\,s^{-1}}$, we obtained for the mixing timescale within the SB cavity, $\tau_{\text{m}}\sim \ell/a$, a value of about $\unit[1]{Myr}$. Considering that the last SN occurred roughly $\unit[1.5]{Myr}$ ago, it is therefore not surprising that the \element[][60]{Fe} mass is distributed quite uniformly in the LB cavity at present time.
%

\subsection{Comparison with crust measurements}
\label{ch:compa}
\begin{figure}
   \resizebox{\hsize}{!}{\includegraphics{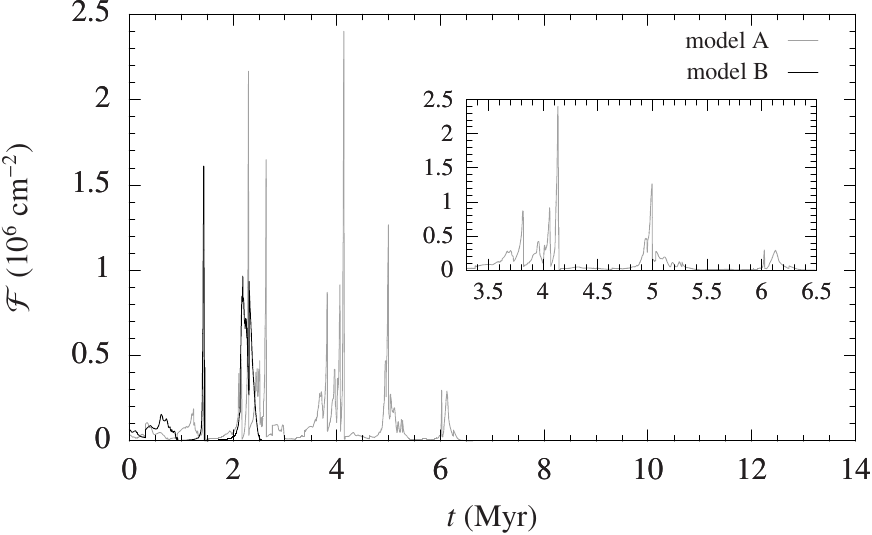}}
      \caption{Local interstellar fluence of \element[][60]{Fe} atoms due to all SNe in the LB volume as a function of time before present in model A (grey line) and B (black line). The inlay magnifies the three different enhancement types, being SN blast waves (high and narrow signals), reflected shocks (weaker and broader signals), and the supershell (very broad rightmost signal).
              }
         \label{im:fluence_tot_AB}
   \end{figure}

At each simulation time step, the total local interstellar fluence of \element[][60]{Fe} atoms was calculated according to Eq.~(\ref{eq:feflu}). Figure \ref{im:fluence_tot_AB} shows the resulting profiles for both models. Upon closer inspection (see inlay), the signals come in three different types, which are all embedded in a `background noise' of $\simeq$$\unit[10^5]{cm^{-2}}$ amplitude.

The first type are sharp and intense ($\mathcal{F}\simeq 0.3$--$\unit[2.4\times 10^6]{cm^{-2}}$) sawtooth waves that occur seven times in model A, but only once in model B. Such a wave form is characteristic of the density profile of an SN remnant in the Sedov-Taylor phase. These signals must hence originate from the shells of individual remnants that cross the Earth's orbit while expanding into the LB volume. The time of direct deposition of the SN ejecta on Earth, given by the (scale) width of the signal, is a measure of the remnant's shell thickness. The time range of direct exposure in our models, $\Delta t\simeq 70$--$\unit[130]{kyr}$, roughly lies between the values proposed by \cite{Fie:05} and \cite{Bis:11}, but almost perfectly matches the predictions given in \cite{Fei:14}, which are based on the SN model by \cite{Kah:98}.

The second type, which appears to be unique to model A, are signals that occur after almost every sawtooth wave (note that the time axis is from right to left), while being always more extended and weaker than the waves they succeed, almost as if they were the SN explosions' `echoes'. Remarkably, the time lag between the two signals increases the closer the present time is approached.

The third type is a joint feature of both profiles, occurring only once and right at the beginning ($t\simeq 6.1$ and $\unit[2.2]{Myr}$ in model A and B, respectively). Here, the signals are quite broad ($\Delta t\simeq 300$ and $\unit[450]{kyr}$ in model A and B, respectively) and possess intermediate amplitudes of about $2.9\times 10^5$ (model A) and $\unit[9.6\times 10^5]{cm^{-2}}$ (model B).

   \begin{figure}
   \resizebox{\hsize}{!}{\includegraphics{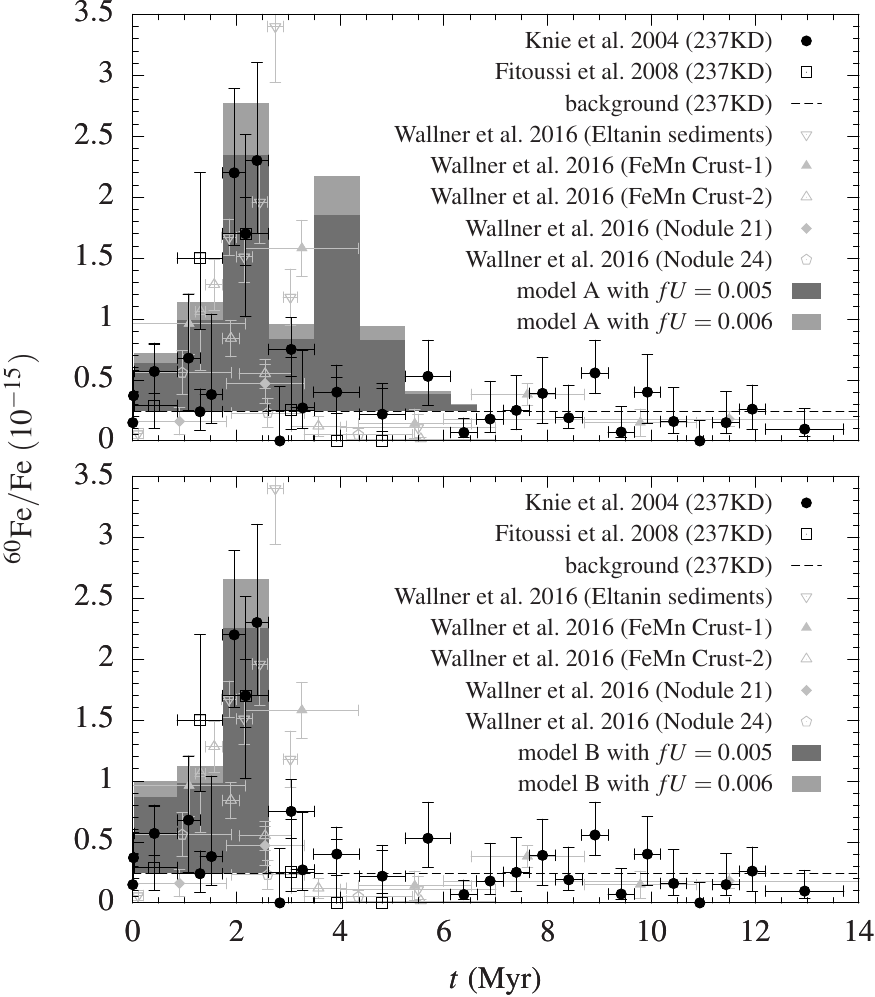}}
      \caption{Comparison of the measured (symbols with error bars) and simulated (histograms) \element[][60]{Fe}/Fe ratio as a function of the terrestrial archive's layer age for model A (\emph{upper panel}) and B (\emph{lower panel}). Simulation data is set upon the instrumental background derived specifically for 237KD, which is indicated by the dashed line. The grey levels of the histograms refer to the \element[][60]{Fe} survival fractions employed.}
         \label{im:vglcrust_AB}
   \end{figure}

Applying the rebinning procedure described in Sect.~\ref{ch:fe60_calc} allowed us to do a direct comparison with the FeMn crust measurements of \cite{Kni:04} and \cite{Fit:08}, as shown in Fig.~\ref{im:vglcrust_AB}. The \element[][60]{Fe} survival fraction, $fU$, was either set to $0.006$ (light grey histograms), as originally suggested by \citeauthor{Kni:04}, or 0.005 (dark grey histograms), corresponding to a combination of a dust fraction of $f=0.01$ with the newly derived lower limit for $U$ (cf.~Sect.~\ref{ch:fe60_calc}). It is seen that for both models, measurements are best matched if the slightly lower uptake factor is employed. This is particularly true for the (global) maximum in the crust layer corresponding to $t\simeq \unit[2.2]{Myr}$, whose presence was lately confirmed by the study of \citet[see grey symbols in Fig.~\ref{im:vglcrust_AB}]{Wal:16}, and which is perfectly reproduced by both models ($\max (^{60}\text{Fe}/\text{Fe})\simeq 2.3\times 10^{-15}$). It is also interesting to note that model A suggests contributions in five more layers than model B, with two of those, namely for $t\simeq 3.1$ and $\unit[5.7]{Myr}$, lying within the error bars of the measurements. The local maximum calculated for $t\simeq\unit[3.9]{Myr}$ is however absent in the FeMn crust data available.

   \begin{figure}
   \resizebox{\hsize}{!}{\includegraphics{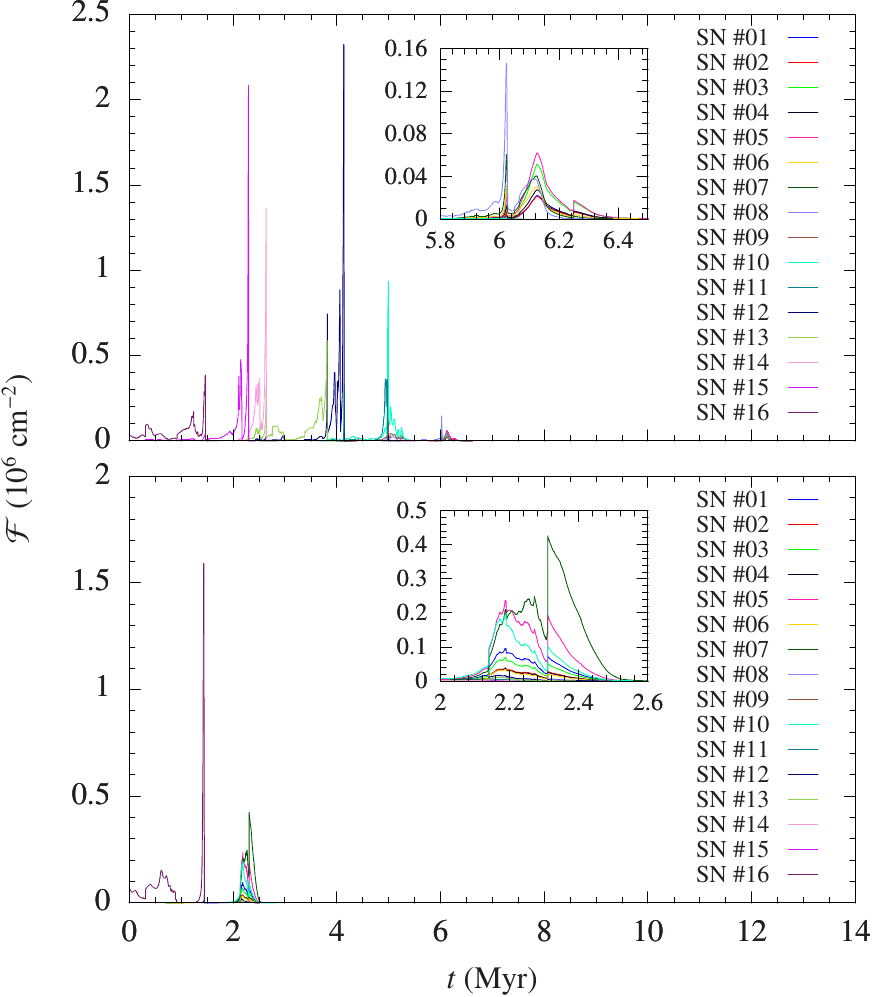}}
      \caption{Local interstellar fluence of \element[][60]{Fe} atoms due to each individual SN in the LB volume (see legends) as a function of time before present in model A (\emph{upper panel}) and B (\emph{lower panel}). The inlays are zoom-ins on the last group of peaks, respectively.
              }
         \label{im:fluence_AB}
   \end{figure}

To explore the exact origin of each signal in the fluence profiles, we rerun the simulations with all parameters unchanged, except that we assigned a passive scalar to each individual SN event. Formally, this implies adding an index $i$ with $1\le i\le 16$ to the variables $\mathcal{F}$, $Z$, and $\Sigma$ in Eqs.~(\ref{eq:feflu}) and (\ref{eq:fesurfden}). Results are plotted in Fig.~\ref{im:fluence_AB}, where identification numbers correspond to the explosion order. Inspecting these profiles, together with the left column panels of Figs.~\ref{im:fedmaps_A} and \ref{im:fedmaps_B}, reveals that the earliest signals of both models, i.e.~the quite broad peaks of intermediate amplitude already known from Fig.~\ref{im:fluence_tot_AB}, are produced when the LB's supershell runs over the solar system. Separated from the shocked ambient medium by a contact discontinuity, the shell harbours a large amount of \element[][60]{Fe} atoms that have by then been expelled in previous SN events (SNe \#01--08 and \#01--15 in model A and B, respectively) and that have not yet decayed (see also the inlays of Fig.~\ref{im:fluence_AB}). Since the supershell of model B arrives later at the Earth's orbit, it could already accumulate more SN ejecta (and thus enriched material), which is the reason why the profile of the total local interstellar fluence (Fig.~\ref{im:fluence_tot_AB}) has a first signal that is both broader and higher than in model A. Also note that the first signals do not set in abruptly but in a smooth, ramp-like manner. This is due to the combined effects of numerical diffusion present in all Godunov-like advection schemes and the Rayleigh-Taylor instability operating at the contact discontinuity inside the supershell (see Sect.~\ref{ch:evo_AB}), allowing for mixing across the discontinuity. 

Even the pulses generated by single SN remnants actually entrain fractions of previously released \element[][60]{Fe}, since the average time between successive explosions is short when compared to the half-life of \element[][60]{Fe}. This is nicely confirmed by Fig.~\ref{im:fluence_AB}, showing that the high peaks associated with recent SN events indeed superpose a series of peaks with lower amplitudes. The same can be observed for the weaker follow-up signals, which are produced when the SN blast waves return after being reflected from the surrounding supershell; so in this sense they are indeed SN `echoes'. Finally, since the supershell moves away from Earth, the time lag between an explosion and its `echo' becomes greater for later times.

Putting all this information together, we arrive at the following interpretation of Fig.~\ref{im:vglcrust_AB}. Starting with the most prominent feature, the (global) maximum in the layer for $t\simeq\unit[2.2]{Myr}$, we can now safely state that, in model A, it is produced by SN \#14 and 15 and their reflected shocks, whereas in model B, it is due to yet undecayed material from SNe \#01--15 reaching Earth almost simultaneously while being gathered within the LB's supershell (see also the middle and left column panels of Figs.~\ref{im:fedmaps_A} and \ref{im:fedmaps_B}, respectively). The additional local maximum at $t\simeq\unit[3.9]{Myr}$, which is unique to model A, is mainly due to SN \#12 and 13. The upper crust layers (corresponding to $t \lesssim\unit[1.7]{Myr}$) obtain their \element[][60]{Fe} content almost exclusively from SN \#16. There is at first the arrival of the SN blast wave itself (seen for model B in the middle column panel of Fig.~\ref{im:fedmaps_B}) that generates the enhancement in the layer for $t\simeq \unit[1.3]{Myr}$\footnote{Note that in model B, the corresponding peak also contains the contribution of the reflected shock, which could already return due to the supershell's proximity, making clear why the signal has such a high amplitude. In contrast, there is a time lag of about $\unit[230]{kyr}$ in model A.}. Later on, turbulent motions induced by the blast wave of the last SN lead to the enhancement in the layer for $t\simeq\unit[0.4]{Myr}$. Particularly, \emph{all} the `background noise' in the modelled fluence profiles is due to turbulence inside the LB cavity. 

We hence conclude that the measured \element[][60]{Fe} excess in the FeMn crust could be either due to the passages of dust-loaded shells from nearby SN remnants over the Earth's orbit, sweeping up also enriched material, which was present at that time in the LB volume (model A), or, alternatively, due to the arrival of the supershell of the LB that incorporates the material from several previous SN events (model B). We consider the LB formation scenario of model B to be physically more probable than that of A. Not only are the derived and measured $^{60}\text{Fe}/\text{Fe}$ ratios in better agreement (there is particularly no second maximum), but also the size and (more importantly for simulations that are based on homogeneous background media) the gas-dynamical properties of the fully developed SB are a better match to observations.

We further emphasise that the additional smaller peak at $t\simeq6.5$--$\unit[8.7]{Myr}$, partly present in the data of \citet[filled triangles in Fig.~\ref{im:vglcrust_AB}]{Wal:16}, does not occur in our profiles. If this enhancement is not of meteoritic origin, as mentioned earlier, but actually connected to the formation of the LB, it would be an indication for a clustering of stars between 12 and $\unit[15]{M_{\sun}}$, thus differing from the most probable binning used for the present study. However, this hypothesis needs further testing and simulations, and is beyond the scope of this work, but will be addressed in a forthcoming paper.
%

\section{Discussion and conclusions}
\label{ch:concl}
In this paper we presented three-dimensional HD simulations (with resolutions adaptively adjusted down to subparsec scale) aiming to reproduce an \element[][60]{Fe} excess measured in a deep-sea FeMn crust in the context of the formation of the LB. Parameters for the LB-generating SNe are based on an IMF normalised to still existing low mass stars of the parent stellar moving group identified by \cite{Fuc:06}. Like in the analytical model by \cite{Fei:10} (see also \citetalias{Bre:16}), ejected masses, lifetimes and thus explosion times were derived from the initial masses of the progenitor stars, whereas computations of the most probable paths for the individual perished members of the moving group provided most probable explosion sites. For estimating the particular \element[][60]{Fe} mass ejecta, we resorted to predictions from stellar evolution models. The formation of the LB, as well as of the neighbouring Loop~I SB, was studied in two homogeneous, self-gravitating environments, sharing similarities with the classical warm ionised (model A) and warm neutral medium (model B). Interstellar heating and cooling processes were included via standard CIE net cooling functions. By utilising passive scalars for tagging the \element[][60]{Fe} enriched gas released into the SB cavities, we recorded the flux of \element[][60]{Fe} atoms at the Earth's orbit (usually termed local interstellar fluence) over the whole simulation time, and then calculated time averages over a range matching the crust layers in order to do comparisons with the measurements of \cite{Kni:04} and \cite{Fit:08}. Our models are able to reproduce both the observed timing and intensity of the \element[][60]{Fe} excess with rather high precision. However, the underlying physical processes are different. In model A the signal arises due to the fast-paced blast waves of two individual SN remnants (\#14 and 15 out of 16 in total), which cross the Earth's orbit twice as a result of reflection from the LB's outer shell. In model B, on the other hand, it is the supershell of the LB itself that injects the yet undecayed \element[][60]{Fe} content of all previous SNe (\#01--15 in this case) at once, but over a longer time range. For both scenarios it is important to note that \element[][60]{Fe} does not reach Earth in gaseous form, but condensed onto dust grains that have to survive a variety of filtering processes throughout their journey. These were summarised by us into the combined multiplicative factor $fU$, for which we found a better agreement with the crust data if $fU=0.005$ rather than 0.006. Both values occur in the current literature. The observed extension of the Local Bubble as well as its gas-dynamical properties were better matched by model B. The age of the LB has been estimated from the main sequence lifetime of a still existing $\unit[8.2]{M_{\sun}}$ star \citep[cf.][]{Fuc:06}.


Primarily for the sake of better comparison with analytical calculations (\citealt{Fei:10}, \citetalias{Bre:16}) our model does currently not include the stellar winds blown by the moving group members. For a locally homogeneous ISM, as considered in this work, these winds would have carved an extended, egg-like shaped cavity around the cluster's center-of-mass trajectory, already before the first stars explode \citep[cf.][and references therein]{Wea:77}. As soon as this happens, however, the dynamics of the wind-blown bubble is quickly overtaken by the SNe, as is clear from comparing the mechanical luminosities of the winds and SNe \citep[cf.][]{Tom:81}. The former cannot be larger than $\frac{1}{2}\dot{M}_\mathrm{tot}v_\infty^2\simeq\unit[2.5\times 10^{36}]{erg\,s^{-1}}$, where $\dot{M}_\mathrm{tot}\simeq\unit[2\times10^{-6}]{M_\odot\,yr^{-1}}$ is the maximum stellar mass loss rate (only reached if the most massive star is still alive) and $v_\infty\simeq\unit[2000]{km\,s^{-1}}$ is the terminal wind velocity. The latter, on the other hand, is $E_\mathrm{SN}/\Delta\tau\simeq\unit[4.2\times 10^{37}]{erg\,s^{-1}}$, with $\Delta\tau\simeq\unit[7.4\times 10^5]{yr}$ being the average time interval between two consecutive SNe. Note that we only consider \emph{cumulative} mechanical energy output rates here, which should be justified because the mean separation between the massive stars (and hence the SN explosion centres) of $\simeq$$\unit[44]{pc}$ is most of the time smaller than the size of the SB. So the expansion of mature SBs is determined by SNe, rather than by stellar winds (and also ionising photons), which dominate the early phase of evolution \citep[see also][]{Rog:13,Kra:13,Yad:16,Kim:16,Kim:17}.

Although our simulations draw a quite conclusive picture of the LB formation scenario, there are nevertheless some caveats that have to be addressed. 

First, the production and expulsion of \element[][60]{Fe} is a complicated process that strongly depends on the mixing processes in the interiors of massive stars, and the depth of their convection zones. Predictions of the \element[][60]{Fe} yields can thus vary by a factor of a few, even for the same stellar masses, depending on which stellar evolution model is used \citep[cf.][]{Tur:10}.

Second, the uptake factor, quantifying the sedimentation through the Earth's atmosphere onto the ocean floor, is only loosely constrained. More precise estimates, either from measuring \element[][60]{Fe} at other sites or from the analysis of other radioisotopes, are therefore highly welcome, since its value has a profound effect on our quantative results.

The third concerns heliospheric filtering. \element[][60]{Fe}-enriched plasma could only reach the solar system via the poles because of strong ram pressure of the solar wind in the equatorial plane. Due to its higher mass this does not affect \element[][60]{Fe}-dust. A comprehensive description of the formation and survival of SN dust at different compositions and sizes is therefore desirable, as it is particularly questionable whether the single factor applied in this work can fully account for all processes involved. In principle, similar arguments apply for the previous two points. An obvious advantage of employing simple input parameters however lies in the fact that they can be easily adapted when the underlying physics is better understood or measurements have become more precise. Still, it should be emphasised that these uncertainties affect the absolute values of \element[][60]{Fe}, but not relative ones, giving the model a solid foundation.

Last but not least, there is the dependence on the background medium, as well as on the explosion sites in the stellar moving group. For the former we can say from our current studies with a homogeneous background medium that the average external density may not exceed $\simeq$$\unit[0.3]{cm^{-3}}$. Otherwise, the LB supershell would arrive later and the \element[][60]{Fe} overabundance would no longer appear in the crust layer in which it was discovered. Yet strictly speaking, this is only true for the SN explosion centres applied in this work, which we left unchanged as these are the most probable ones for the considered stellar moving group (see also \citetalias{Bre:16}). One can however check whether any relevant changes in the calculated \element[][60]{Fe}/Fe profiles (Fig.~\ref{im:vglcrust_AB}) would result if an SN would occur by a maximum of 1$\sigma$ away from its most probable explosion site lying at distance $d$. In order to estimate this spread in distance, $\delta R$, we traced back the paths of 1000 hypothetical stars that would explode on exactly the same location $\unit[10]{Myr}$ into the past, while taking into account the error in their mean velocities ($\simeq$$\unit[3]{km\,s^{-1}}$ for each component). We obtained $\delta R\simeq\unit[20]{pc}$. As the blast wave velocity in the LB volume is of the order of $\unit[10^4]{km\,s^{-1}}$, this translates to a spread in the arrival time of a thus generated \element[][60]{Fe} signal of $\simeq$$\unit[2]{kyr}$, which lies far below the average time resolution of the FeMn crust sample ($\simeq$$\unit[0.6]{Myr}$, see the bin sizes in Fig.~\ref{im:vglcrust_AB}). Relevant for the height of the signal is the size of the SN shells when they are crossing our solar system. When assuming for the sake of simplicity perfect sphericity and neglecting the clumpiness of these shells as a result of instabilities, the factor by which the height of the signal would change lies between $(1\pm \delta R/d)^2$. For the last three SNe in the LB volume $d\simeq \unit[100]{pc}$, yielding for the factor's lower and upper limit  0.64 and 1.44, respectively.

Directly related to the background medium is the question about the amount of \element[][60]{Fe} in the region where the LB began to form. The steady-state Galactic \element[][60]{Fe} mass of $\unit[1.53]{M_\odot}$ given by \cite{Die:16} was derived for the diffuse, extended ISM and thus does not allow conclusions to be drawn on local values in the solar neighborhood today, and even less on those of more than $\unit[10]{Myr}$ ago, where also radioactive decay should play a crucial role. The only serious way to obtain such information would be long-term chemical evolution simulations of the (local) Galaxy that particularly include chemical mixing. As such models are currently unavailable, we set in our simulations the initial amount of \element[][60]{Fe} equal to zero everywhere, thus always representing the lower limit case. We note in passing that a completely analogous problem arises in the context of explaining the origin of \element[][60]{Fe} in the early solar system \citep[see e.g.][]{Was:98,Hus:09,Bos:13,Vas:13}, demonstrating that such studies are strongly needed.

Also worth mentioning is the alternative explanation for the $\unit[2.2]{Myr}$-old \element[][60]{Fe} signal given by \cite{Bas:07} and \cite{Stu:12}. Their idea is based on the fact that \element[][60]{Fe} is not only produced in SNe but also by spallation reactions between Galactic cosmic rays and nickel in \mbox{(micro-)} meteorites and interplanetary dust particles. Via collisions of asteroids, e.g.~in the main belt, these small objects could have been released in large amounts, thus exposing Earth to a boosted isotopic influx. However, the concentration of nickel measured within the FeMn crust is too low to explain the \element[][60]{Fe} signal within the framework of this scenario \citep{Fit:08,Wal:16}. On the other hand, the cosmic-ray protons, antiprotons, and positrons that would have been copiously released by a nearby SN would certainly have led to a temporal increase of radioisotopes in meteoritic particles. As a matter of fact, there are evidences for one (or more) such events about $\unit[2]{Myr}$ ago in the cosmic ray energy spectra \citep[e.g.][]{Kac:15}, fitting nicely into the picture drawn in this work.

Another strong hint for the SN origin of the \element[][60]{Fe} signal are the experimental \element[][60]{Fe} and \element[][53]{Mn} activities recently obtained from lunar samples \citep{Fim:16}. Like \element[][60]{Fe}, the long-lived radionuclide \element[][53]{Mn} is generated from cosmic-ray spallation reactions on iron in meteorites. Hence, if the \element[][60]{Fe} signal was produced from an enhanced influx of micrometeorites, one would expect the lunar activity ratios of \element[][60]{Fe}/\element[][53]{Mn} to be consistent with meteoritic values. Yet the lunar soil samples carry an enhanced \element[][60]{Fe} signature that exceeds the activity created by cosmic rays.

We can actually test our numerical calculations against these lunar soil measurements, stating an observed fluence in the range between $10^7$ and $\unit[6\times 10^7]{cm^{-2}}$. As their data is not time-resolved, we have to sum the value of $\Sigma$ (cf.~Eq.~(\ref{eq:fesurfden})) in all time bins available, spanning a total range of $\unit[12.6]{Myr}$. When setting $U=f=1$, we obtain fluences of $\unit[6.18\times 10^8]{cm^{-2}}$ for model A and $\unit[3.36\times 10^8]{cm^{-2}}$ for model B. While it is almost certain that $f<1$ (cf.~Sect.~\ref{ch:fe60_calc}), the Moon's uptake factor should be, due to the lack of an atmosphere, equal or very close to unity. Taking our simulations as a basis, we estimate a range of 0.016 and 0.179 for the \element[][60]{Fe} survival fraction $fU$. The lower limit would be actually compatible to an uptake of 100\,\%, if the value of $f$ were only slightly higher than estimated by \cite{Fry:15}, namely 0.016 instead of 0.010. Therefore also the measurements of \element[][60]{Fe} in lunar samples are in agreement with our previously performed simulations.

\begin{acknowledgements}
M.M.S. and D.B. acknowledge funding by the DFG priority program 1573 ``Physics of the Interstellar Medium''. We thank the anonymous referee for her/his valuable comments which helped to improve the manuscript.
\end{acknowledgements}

\bibliographystyle{aa}
\bibliography{LBI}

\begin{thebibliography}{105}
\expandafter\ifx\csname natexlab\endcsname\relax\def\natexlab#1{#1}\fi

\bibitem[{Balsara {et~al.}(2004)Balsara, Kim, {Mac Low}, \& Mathews}]{Bal:04}
Balsara, D.~S., Kim, J., {Mac Low}, M., \& Mathews, G.~J. 2004, Astrophys. J.,
  617, 339

\bibitem[{Bania \& Lyon(1980)}]{Ban:80}
Bania, T.~M. \& Lyon, J.~G. 1980, Astrophys. J., 239, 173

\bibitem[{Basu {et~al.}(2007)Basu, Stuart, Schnabel, \& Klemm}]{Bas:07}
Basu, S., Stuart, F.~M., Schnabel, C., \& Klemm, V. 2007, Phys. Rev. Lett., 98,
  141103

\bibitem[{Bergh{\"{o}}fer \& Breitschwerdt(2002)}]{Ber:02}
Bergh{\"{o}}fer, T.~W. \& Breitschwerdt, D. 2002, Astron. Astrophys., 390, 299

\bibitem[{Binns {et~al.}(2016)Binns, Israel, Christian, Cummings, de~Nolfo,
  Lave, Leske, Mewaldt, Stone, von Rosenvinge, \& Wiedenbeck}]{Bin:16}
Binns, W.~R., Israel, M.~H., Christian, E.~R., {et~al.} 2016, Science, 352, 677

\bibitem[{Bishop \& Egli(2011)}]{Bis:11}
Bishop, S. \& Egli, R. 2011, Icarus

\bibitem[{Bizzarro {et~al.}(2007)Bizzarro, Ulfbeck, Trinquier, Thrane,
  Connelly, \& Meyer}]{Biz:07}
Bizzarro, M., Ulfbeck, D., Trinquier, A., {et~al.} 2007, Sci, 316, 1178

\bibitem[{Boss \& Keiser(2013)}]{Bos:13}
Boss, A.~P. \& Keiser, S.~A. 2013, Astrophys. J., 770, 51

\bibitem[{Breitschwerdt \& de~Avillez(2006)}]{Bre:06}
Breitschwerdt, D. \& de~Avillez, M.~A. 2006, Astron. Astrophys., 452, L1

\bibitem[{Breitschwerdt {et~al.}(2016)Breitschwerdt, Feige, Schulreich, de.
  Avillez, Dettbarn, \& Fuchs}]{Bre:16}
Breitschwerdt, D., Feige, J., Schulreich, M.~M., {et~al.} 2016, Nature, 532, 73

\bibitem[{Breitschwerdt {et~al.}(2000)Breitschwerdt, Freyberg, \&
  Egger}]{Bre:00}
Breitschwerdt, D., Freyberg, M.~J., \& Egger, R. 2000, Astron. Astrophys., 361,
  303

\bibitem[{Busso {et~al.}(2003)Busso, Gallino, \& Wasserburg}]{Bus:03}
Busso, M., Gallino, R., \& Wasserburg, G.~J. 2003, Publ. Astron. Soc. Aust.,
  20, 356

\bibitem[{Chiang \& Bregman(1988)}]{Chi:88}
Chiang, W.-H. \& Bregman, J.~N. 1988, Astrophys. J., 328, 427

\bibitem[{Chiang \& Prendergast(1985)}]{Chi:85}
Chiang, W.-H. \& Prendergast, K.~H. 1985, Astrophys. J., 297, 507

\bibitem[{Davidson(2004)}]{Dav:04}
Davidson, P.~A. 2004, {Turbulence: An Introduction for Scientists and
  Engineers}, 1st edn. (Oxford: Oxford University Press), 234--235

\bibitem[{de~Avillez \& Breitschwerdt(2004)}]{Avi:04}
de~Avillez, M.~A. \& Breitschwerdt, D. 2004, Astron. Astrophys., 425, 899

\bibitem[{de~Avillez \& Breitschwerdt(2005)}]{Avi:05}
de~Avillez, M.~A. \& Breitschwerdt, D. 2005, Astron. Astrophys., 436, 585

\bibitem[{de~Avillez \& Breitschwerdt(2009)}]{Avi:09}
de~Avillez, M.~A. \& Breitschwerdt, D. 2009, Astrophys. J., 697, L158

\bibitem[{de~Avillez \& Breitschwerdt(2012)}]{Avi:12}
de~Avillez, M.~A. \& Breitschwerdt, D. 2012, Astron. Astrophys., 539, L1

\bibitem[{de~Avillez \& {Mac Low}(2002)}]{Avi:02}
de~Avillez, M.~A. \& {Mac Low}, M.-M. 2002, Astrophys. J., 581, 1047

\bibitem[{Diehl(2016)}]{Die:16}
Diehl, R. 2016, J. Phys.: Conf. Ser., 665, 012011

\bibitem[{Dobbs {et~al.}(2011)Dobbs, Burkert, \& Pringle}]{Dob:11}
Dobbs, C.~L., Burkert, A., \& Pringle, J.~E. 2011, Mon. Not. R. Astron. Soc.,
  417, 1318

\bibitem[{Egger \& Aschenbach(1995)}]{Egg:95}
Egger, R.~J. \& Aschenbach, B. 1995, Astron. Astrophys., 294, L25

\bibitem[{Ellis {et~al.}(1996)Ellis, Fields, \& Schramm}]{Ell:96}
Ellis, J., Fields, B.~D., \& Schramm, D.~N. 1996, Astrophys. J., 470, 1227

\bibitem[{Farley {et~al.}(2006)Farley, Vokrouhlick{\'{y}}, Bottke, \&
  Nesvorn{\'{y}}}]{Far:06}
Farley, K.~A., Vokrouhlick{\'{y}}, D., Bottke, W.~F., \& Nesvorn{\'{y}}, D.
  2006, Nature, 439, 295

\bibitem[{Feige(2010)}]{Fei:10}
Feige, J. 2010, Master's thesis, University of Vienna

\bibitem[{Feige(2014)}]{Fei:14}
Feige, J. 2014, PhD thesis, University of Vienna

\bibitem[{Feige {et~al.}(2013)Feige, Wallner, Fifield, Korschinek, Merchel,
  Rugel, Steier, Winkler, \& Golser}]{Fei:13}
Feige, J., Wallner, A., Fifield, L., {et~al.} 2013, EPJ web conf., 63, 03003

\bibitem[{Feige {et~al.}(2012)Feige, Wallner, Winkler, Merchel, Fifield,
  Korschinek, Rugel, \& Breitschwerdt}]{Fei:12}
Feige, J., Wallner, A., Winkler, S.~R., {et~al.} 2012, Publ. Astron. Soc.
  Aust., 29, 109

\bibitem[{Ferland {et~al.}(1998)Ferland, Korista, Verner, Ferguson, Kingdon, \&
  Verner}]{Ferl:98}
Ferland, G.~J., Korista, K.~T., Verner, D.~A., {et~al.} 1998, Publ. Astron.
  Soc. Pac., 110, 761

\bibitem[{Fields \& Ellis(1999)}]{Fie:99}
Fields, B.~D. \& Ellis, J. 1999, New Astron., 4, 419

\bibitem[{Fields {et~al.}(2005)Fields, Hochmuth, \& Ellis}]{Fie:05}
Fields, B.~D., Hochmuth, K.~A., \& Ellis, J. 2005, Astrophys. J., 621, 902

\bibitem[{Fimiani {et~al.}(2016)Fimiani, Cook, Faestermann,
  G{\'{o}}mez-Guzm{\'{a}}n, Hain, Herzog, Knie, Korschinek, Ludwig, Park,
  Reedy, \& Rugel}]{Fim:16}
Fimiani, L., Cook, D.~L., Faestermann, T., {et~al.} 2016, Phys. Rev. Lett., 116

\bibitem[{Fitoussi {et~al.}(2008)Fitoussi, Raisbeck, Knie, Korschinek,
  Faestermann, Goriely, Lunney, Poutivtsev, Rugel, Waelbroeck, \&
  Wallner}]{Fit:08}
Fitoussi, C., Raisbeck, G., Knie, K., {et~al.} 2008, Phys. Rev. Lett., 101,
  121101

\bibitem[{Fry {et~al.}(2015)Fry, Fields, \& Ellis}]{Fry:15}
Fry, B.~J., Fields, B.~D., \& Ellis, J.~R. 2015, Astrophys. J., 800, 71

\bibitem[{Fry {et~al.}(2016)Fry, Fields, \& Ellis}]{Fry:16}
Fry, B.~J., Fields, B.~D., \& Ellis, J.~R. 2016, Astrophys. J., 827, 48

\bibitem[{Fuchs {et~al.}(2006)Fuchs, Breitschwerdt, de~Avillez, Dettbarn, \&
  Flynn}]{Fuc:06}
Fuchs, B., Breitschwerdt, D., de~Avillez, M.~A., Dettbarn, C., \& Flynn, C.
  2006, Mon. Not. R. Astron. Soc., 373, 993

\bibitem[{Galeazzi {et~al.}(2014)Galeazzi, Chiao, Collier, Cravens, Koutroumpa,
  Kuntz, Lallement, Lepri, McCammon, Morgan, Porter, Robertson, Snowden,
  Thomas, Uprety, Ursino, \& Walsh}]{Gal:14}
Galeazzi, M., Chiao, M., Collier, M.~R., {et~al.} 2014, Nature, 512, 171

\bibitem[{Girichidis {et~al.}(2015)Girichidis, Naab, Walch, Hanasz, {Mac Low},
  Ostriker, Gatto, Peters, W{\"{u}}nsch, Glover, Klessen, Clark, \&
  Baczynski}]{Gir:15}
Girichidis, P., Naab, T., Walch, S., {et~al.} 2015, Astrophys. J. Lett., 816

\bibitem[{Gressel {et~al.}(2008)Gressel, Elstner, Ziegler, \&
  R{\"{u}}diger}]{Gre:08a}
Gressel, O., Elstner, D., Ziegler, U., \& R{\"{u}}diger, G. 2008, Astron.
  Astrophys., 486, L35

\bibitem[{Hanasz {et~al.}(2009)Hanasz, W{\'{o}}ltanski, Kowalik, \&
  Pawlaszek}]{Han:09}
Hanasz, M., W{\'{o}}ltanski, D., Kowalik, K., \& Pawlaszek, R. 2009, Astrophys.
  J., 706, L155

\bibitem[{Huss {et~al.}(2009)Huss, Meyer, Srinivasan, Goswami, \&
  Sahijpal}]{Hus:09}
Huss, G.~R., Meyer, B.~S., Srinivasan, G., Goswami, J.~N., \& Sahijpal, S.
  2009, Geochim. Cosmochim. Acta, 73, 4922

\bibitem[{Joung {et~al.}(2009)Joung, {Mac Low}, \& Bryan}]{Jou:09}
Joung, M.~R., {Mac Low}, M.-M., \& Bryan, G.~L. 2009, Astrophys. J., 704, 137

\bibitem[{Kachelrie{\ss} {et~al.}(2015)Kachelrie{\ss}, Neronov, \&
  Semikoz}]{Kac:15}
Kachelrie{\ss}, M., Neronov, A., \& Semikoz, D.~V. 2015, Phys. Rev. Lett., 115,
  181103

\bibitem[{Kahn(1998)}]{Kah:98}
Kahn, F.~D. 1998, Lect. Notes Phys., 506, 483

\bibitem[{Kim \& Ostriker(2016)}]{Kim:16}
Kim, C.-G. \& Ostriker, E.~C. 2016, ArXiv e-prints [\eprint[arXiv]{1612.03918}]

\bibitem[{Kim {et~al.}(2017)Kim, Ostriker, \& Raileanu}]{Kim:17}
Kim, C.-G., Ostriker, E.~C., \& Raileanu, R. 2017, Astrophys. J., 834, 25

\bibitem[{Knie {et~al.}(2004)Knie, Korschinek, Faestermann, Dorfi, Rugel, \&
  Wallner}]{Kni:04}
Knie, K., Korschinek, G., Faestermann, T., {et~al.} 2004, Phys. Rev. Lett., 93,
  171103

\bibitem[{Knie {et~al.}(1999)Knie, Korschinek, Faestermann, Wallner, Scholten,
  \& Hillebrandt}]{Kni:99}
Knie, K., Korschinek, G., Faestermann, T., {et~al.} 1999, Phys. Rev. Lett., 83,
  18

\bibitem[{Korschinek {et~al.}(1996)Korschinek, Faestermann, Knie, \&
  Schmidt}]{Kor:96}
Korschinek, G., Faestermann, T., Knie, K., \& Schmidt, C. 1996, Radiocarbon,
  38, 68

\bibitem[{Krause {et~al.}(2013)Krause, Fierlinger, Diehl, Burkert, Voss, \&
  Ziegler}]{Kra:13}
Krause, M., Fierlinger, K., Diehl, R., {et~al.} 2013, Astron. Astrophys., 550,
  A49

\bibitem[{Lallement {et~al.}(2014)Lallement, Vergely, Valette, Puspitarini,
  Eyer, \& Casagrande}]{Lal:14}
Lallement, R., Vergely, J.-L., Valette, B., {et~al.} 2014, Astron. Astrophys.,
  561, A91

\bibitem[{Lallement {et~al.}(2003)Lallement, Welsh, Vergely, Crifo, \&
  Sfeir}]{Lal:03}
Lallement, R., Welsh, B.~Y., Vergely, J.~L., Crifo, F., \& Sfeir, D. 2003,
  Astron. Astrophys., 411, 447

\bibitem[{Langevin \& Arnold(1977)}]{Lan:77}
Langevin, Y. \& Arnold, J.~R. 1977, Annu. Rev. Earth Planet. Sci., 5, 449

\bibitem[{Lax(1954)}]{Lax:54}
Lax, P.~D. 1954, Comm. Pure Appl. Math., 7, 159

\bibitem[{Limongi \& Chieffi(2006)}]{Lim:06}
Limongi, M. \& Chieffi, A. 2006, Astrophys. J., 647, 483

\bibitem[{Lindblad(1959)}]{Lin:59}
Lindblad, B. 1959, Handb. Phys., 53, 21

\bibitem[{Ludwig {et~al.}(2016)Ludwig, Bishop, Egli, Chernenko, Deneva,
  Faestermann, Famulok, Fimiani, G{\'{o}}mez-Guzm{\'{a}}n, Hain, Korschinek,
  Hanzlik, Merchel, \& Rugel}]{Lud:16}
Ludwig, P., Bishop, S., Egli, R., {et~al.} 2016, Proc. Natl. Acad. Sci. USA,
  113, 9232

\bibitem[{Lugaro {et~al.}(2012)Lugaro, Doherty, Karakas, Maddison, Liffman,
  Garc{\'{i}}a-Hern{\'{a}}ndez, Siess, \& Lattanzio}]{Lug:12}
Lugaro, M., Doherty, C.~L., Karakas, A.~I., {et~al.} 2012, Meteorit. Planet.
  Sci., 47, 1998

\bibitem[{{Mac Low} {et~al.}(1998){Mac Low}, Klessen, Burkert, \&
  Smith}]{Low:98}
{Mac Low}, M.-M., Klessen, R., Burkert, A., \& Smith, M. 1998, Phys. Rev.
  Lett., 80, 2754

\bibitem[{Ma{\'{i}}z-Apell{\'{a}}niz(2001)}]{Mai:01}
Ma{\'{i}}z-Apell{\'{a}}niz, J. 2001, Astrophys. J., 563, 151

\bibitem[{{Ma{\'{i}}z Apell{\'{a}}niz} \& {\'{U}}beda(2005)}]{Mai:05}
{Ma{\'{i}}z Apell{\'{a}}niz}, J. \& {\'{U}}beda, L. 2005, Astrophys. J., 629,
  873

\bibitem[{Massey {et~al.}(1995)Massey, Johnson, \& DeGioia-Eastwood}]{Mas:95}
Massey, P., Johnson, K.~E., \& DeGioia-Eastwood, K. 1995, Astrophys. J., 454,
  151

\bibitem[{McKee \& Ostriker(1977)}]{Kee:77}
McKee, C.~F. \& Ostriker, J.~P. 1977, Astrophys. J., 218, 148

\bibitem[{Mostefaoui {et~al.}(2005)Mostefaoui, Lugmair, \& Hoppe}]{Mos:05}
Mostefaoui, S., Lugmair, G.~W., \& Hoppe, P. 2005, Astrophys. J., 625, 271

\bibitem[{Oegerle {et~al.}(2005)Oegerle, Jenkins, Shelton, Bowen, \&
  Chayer}]{Oeg:05}
Oegerle, W.~R., Jenkins, E.~B., Shelton, R.~L., Bowen, D.~V., \& Chayer, P.
  2005, Astrophys. J., 622, 377

\bibitem[{Padoan \& Nordlund(1999)}]{Pad:99}
Padoan, P. \& Nordlund, A. 1999, Astrophys. J., 526, 279

\bibitem[{Passot {et~al.}(1996)Passot, Vazquez-Semadeni, \& Pouquet}]{Pas:96}
Passot, T., Vazquez-Semadeni, E., \& Pouquet, A. 1996, Astrophys. J., 455, 536

\bibitem[{Piontek \& Ostriker(2007)}]{Pio:07}
Piontek, R.~A. \& Ostriker, E.~C. 2007, Astrophys. J., 663, 183

\bibitem[{Puspitarini {et~al.}(2014)Puspitarini, Lallement, Vergely, \&
  Snowden}]{Pus:14}
Puspitarini, L., Lallement, R., Vergely, J.-L., \& Snowden, S.~L. 2014, Astron.
  Astrophys., 566, A13

\bibitem[{Rauscher {et~al.}(2002)Rauscher, Heger, Hoffman, \& Woosley}]{Rau:02}
Rauscher, T., Heger, A., Hoffman, R.~D., \& Woosley, S.~E. 2002, Astrophys. J.,
  576, 323

\bibitem[{Renaud {et~al.}(2013)Renaud, Bournaud, Emsellem, Elmegreen, Teyssier,
  Alves, Chapon, Combes, Dekel, Gabor, Hennebelle, \& Kraljic}]{Ren:13}
Renaud, F., Bournaud, F., Emsellem, E., {et~al.} 2013, Mon. Not. R. Astron.
  Soc., 436, 1836

\bibitem[{Roe(1986)}]{Roe:86}
Roe, P.~L. 1986, Annu. Rev. Fluid Mech., 18, 337

\bibitem[{Rogers \& Pittard(2013)}]{Rog:13}
Rogers, H. \& Pittard, J.~M. 2013, Mon. Not. R. Astron. Soc., 431, 1337

\bibitem[{Rosen \& Bregman(1995)}]{Ros:95}
Rosen, A. \& Bregman, J.~N. 1995, Astrophys. J., 440, 634

\bibitem[{Rosen {et~al.}(1993)Rosen, Bregman, \& Norman}]{Ros:93}
Rosen, A., Bregman, J.~N., \& Norman, M.~L. 1993, Astrophys. J., 413, 137

\bibitem[{Rugel {et~al.}(2009)Rugel, Faestermann, Knie, Korschinek, Poutivtsev,
  Schumann, Kivel, G{\"{u}}nther-Leopold, Weinreich, \& Wohlmuther}]{Rug:09}
Rugel, G., Faestermann, T., Knie, K., {et~al.} 2009, Phys. Rev. Lett., 103

\bibitem[{Rusanov(1961)}]{Rus:61}
Rusanov, V.~V. 1961, J. Comput. Math. Phys. USSR, 1, 267

\bibitem[{Ryu \& Vishniac(1991)}]{ryu:91}
Ryu, D. \& Vishniac, E.~T. 1991, Astrophys. J., 368, 411

\bibitem[{Savage \& Lehner(2006)}]{Sav:06}
Savage, B.~D. \& Lehner, N. 2006, Astrophys. J. Suppl. Ser., 162, 134

\bibitem[{Schaller {et~al.}(1992)Schaller, Schaerer, Meynet, \&
  Maeder}]{Sch:92}
Schaller, G., Schaerer, D., Meynet, G., \& Maeder, A. 1992, Astron. Astrophys.
  Suppl. Ser., 96, 269

\bibitem[{Schulreich(2015)}]{Schu:15}
Schulreich, M.~M. 2015, PhD thesis, Berlin Institute of Technology

\bibitem[{Slyz {et~al.}(2005)Slyz, Devriendt, Bryan, \& Silk}]{Sly:05}
Slyz, A.~D., Devriendt, J. E.~G., Bryan, G., \& Silk, J. 2005, Mon. Not. R.
  Astron. Soc., 356, 737

\bibitem[{Smith \& Cox(2001)}]{Smi:01}
Smith, R.~K. \& Cox, D.~P. 2001, Astron. Astrophys. Suppl. Ser., 134, 283

\bibitem[{Stone {et~al.}(1998)Stone, Ostriker, \& Gammie}]{Sto:98}
Stone, J.~M., Ostriker, E.~C., \& Gammie, C.~F. 1998, Astrophys. J., 508, L99

\bibitem[{Stuart \& Lee(2012)}]{Stu:12}
Stuart, F.~M. \& Lee, M.~R. 2012, Chem. Geol., 322--323, 209

\bibitem[{Tachibana \& Huss(2003)}]{Tac:03}
Tachibana, S. \& Huss, G.~R. 2003, Astrophys. J., 588, L41

\bibitem[{Teyssier(2002)}]{Tey:02}
Teyssier, R. 2002, Astron. Astrophys., 385, 337

\bibitem[{Timmes {et~al.}(1995)Timmes, Woosley, Hartmann, Hoffman, Weaver, \&
  Matteucci}]{Tim:95}
Timmes, F.~X., Woosley, S.~E., Hartmann, D.~H., {et~al.} 1995, Astrophys. J.,
  449, 204

\bibitem[{Tomisaka {et~al.}(1981)Tomisaka, Habe, \& Ikeuchi}]{Tom:81}
Tomisaka, K., Habe, A., \& Ikeuchi, S. 1981, Astrophys. Space Sci., 78, 273

\bibitem[{Toro(2009)}]{Tor:09}
Toro, E.~F. 2009, {Riemann Solvers and Numerical Methods for Fluid Dynamics: A
  Practical Introduction}, 3rd edn. (Berlin etc.: Springer-Verlag), 25--26

\bibitem[{Tur {et~al.}(2010)Tur, Heger, \& Austin}]{Tur:10}
Tur, C., Heger, A., \& Austin, S.~M. 2010, Astrophys. J., 718, 357

\bibitem[{van Leer(1979)}]{Lee:79}
van Leer, B. 1979, J. Comput. Phys., 32, 101

\bibitem[{Vasileiadis {et~al.}(2013)Vasileiadis, Nordlund, \&
  Bizzarro}]{Vas:13}
Vasileiadis, A., Nordlund, {\AA}., \& Bizzarro, M. 2013, Astrophys. J., 769, L8

\bibitem[{Wallner {et~al.}(2015)Wallner, Bichler, Buczak, Dressler, Fifield,
  Schumann, Sterba, Tims, Wallner, \& Kutschera}]{Wal:15}
Wallner, A., Bichler, M., Buczak, K., {et~al.} 2015, Phys. Rev. Lett., 114,
  041101

\bibitem[{Wallner {et~al.}(2016)Wallner, Feige, Kinoshita, Paul, Fifield,
  Golser, Honda, Linnemann, Matsuzaki, Merchel, Rugel, Tims, Steier, Yamagata,
  \& Winkler}]{Wal:16}
Wallner, A., Feige, J., Kinoshita, N., {et~al.} 2016, Nature, 532, 69

\bibitem[{Wanajo {et~al.}(2013)Wanajo, Janka, \& M{\"{u}}ller}]{Wan:13}
Wanajo, S., Janka, H.-T., \& M{\"{u}}ller, B. 2013, Astrophys. J. Lett., 774,
  L6

\bibitem[{Wang {et~al.}(2007)Wang, Harris, Diehl, Halloin, Cordier, Strong,
  Kretschmer, Kn{\"{o}}dlseder, Jean, Lichti, Roques, Schanne, von Kienlin,
  Weidenspointner, \& Wunderer}]{Wan:07}
Wang, W., Harris, M.~J., Diehl, R., {et~al.} 2007, Astron. Astrophys., 469,
  1005

\bibitem[{Wasserburg {et~al.}(1998)Wasserburg, Gallino, \& Busso}]{Was:98}
Wasserburg, G.~J., Gallino, R., \& Busso, M. 1998, Astrophys. J., 500, L189

\bibitem[{Weaver {et~al.}(1977)Weaver, McCray, Castor, Shapiro, \&
  Moore}]{Wea:77}
Weaver, R., McCray, R., Castor, J., Shapiro, P., \& Moore, R. 1977, Astrophys.
  J., 218, 377 (Erratum 1978, 220, 742)

\bibitem[{Wielen(1982)}]{Wie:82}
Wielen, R. 1982, in Landolt-B{\"{o}}rnstein: Numerical Data and Functional
  Relationships in Science and Technology, Group VI, Vol. 2, ed. K.~Schaifers
  \& H.~H. Voigt (Springer), 225--227

\bibitem[{Woosley(1997)}]{Woo:97}
Woosley, S.~E. 1997, Astrophys. J., 476, 801

\bibitem[{Woosley \& Heger(2007)}]{Woo:07}
Woosley, S.~E. \& Heger, A. 2007, Phys. Rep., 442, 269

\bibitem[{Woosley \& Weaver(1995)}]{Woo:95}
Woosley, S.~E. \& Weaver, T.~A. 1995, Astrophys. J. Suppl. Ser., 101, 181

\bibitem[{Yadav {et~al.}(2016)Yadav, Ray, \& Chakraborti}]{Yad:16}
Yadav, N., Ray, A., \& Chakraborti, S. 2016, Mon. Not. R. Astron. Soc., 459,
  595

\end{thebibliography}

\end{document}